\documentclass[superscriptaddress,twocolumn,showpacs,a4paper,
amssymb,amsmath,nobibnotes,aps,prd,
showkeys,
nofootinbib,notitlepage]{revtex4-1}
\usepackage{verbatim}
\usepackage[T1]{fontenc}
\usepackage[utf8]{inputenc}
\usepackage[american]{babel}
\usepackage{epsfig}
\usepackage{booktabs}
\usepackage{multirow}
\usepackage{dcolumn}
\usepackage{amsmath}
\usepackage{mathtools}
\usepackage{amsfonts}
\usepackage{amssymb}
\usepackage{ulem}
\usepackage{epstopdf}
\usepackage{bm}
\usepackage{siunitx}
\usepackage{braket}
\usepackage{enumitem}
\usepackage{soul}
\usepackage[table]{xcolor}
\usepackage{color}
\usepackage{transparent}
\usepackage{pifont}
\usepackage{enumitem}

\definecolor{navyblue}{rgb}{0.0, 0.0, 0.5}
\definecolor{royalblue}{rgb}{0.25, 0.41, 0.88}
\definecolor{cadmiumgreen}{rgb}{0.0, 0.42, 0.24}
\definecolor{blue-violet}{rgb}{0.54, 0.17, 0.89}
\definecolor{darkviolet}{rgb}{0.58, 0.0, 0.83}
\definecolor{orange(colorwheel)}{rgb}{1.0, 0.5, 0.0}

\usepackage{hyperref}
\hypersetup{
    colorlinks=true, 
    linkcolor=royalblue, 
    citecolor=magenta}

\usepackage{booktabs}
\usepackage{multirow}
\usepackage{dcolumn}
\usepackage{colortbl}

\begin{document}

\title{Tightening the reins on non-minimal dark sector physics:  Interacting Dark Energy with dynamical and non-dynamical equation of state}

\author{William Giar\`e}
\email{w.giare@sheffield.ac.uk}
\affiliation{School of Mathematics and Statistics, University of Sheffield, Hounsfield Road, Sheffield S3 7RH, United Kingdom}

\author{Yuejia Zhai}
\email{y.zhai@sheffield.ac.uk}
\affiliation{School of Mathematics and Statistics, University of Sheffield, Hounsfield Road, Sheffield S3 7RH, United Kingdom}

\author{Supriya Pan}
\email{supriya.maths@presiuniv.ac.in}
\affiliation{Department of Mathematics, Presidency University, 86/1 College Street,  Kolkata 700073, India}
\affiliation{Institute of Systems Science, Durban University of Technology, PO Box 1334, Durban 4000, Republic of South Africa}

\author{Eleonora Di Valentino}
\email{e.divalentino@sheffield.ac.uk}
\affiliation{School of Mathematics and Statistics, University of Sheffield, Hounsfield Road, Sheffield S3 7RH, United Kingdom}

\author{Rafael C. Nunes}
\email{rafadcnunes@gmail.com}
\affiliation{Instituto de F\'{i}sica, Universidade Federal do Rio Grande do Sul, 91501-970 Porto Alegre RS, Brazil}
\affiliation{Divis\~{a}o de Astrof\'{i}sica, Instituto Nacional de Pesquisas Espaciais, Avenida dos Astronautas 1758, S\~{a}o Jos\'{e} dos Campos, 12227-010, S\~{a}o Paulo, Brazil}

\author{Carsten van de Bruck}
\email{c.vandebruck@sheffield.ac.uk}
\affiliation{School of Mathematics and Statistics, University of Sheffield, Hounsfield Road, Sheffield S3 7RH, United Kingdom}

\begin{abstract}
\noindent
We present a comprehensive reassessment of the state of Interacting Dark Energy (IDE) cosmology, namely models featuring a non-gravitational interaction between Dark Matter (DM) and Dark Energy (DE). To achieve high generality, we extend the dark sector physics by considering two different scenarios: a non-dynamical DE equation of state $w_0\neq-1$, and a dynamical $w(a)=w_0+w_a(1-a)$. In both cases, we distinguish two different physical regimes resulting from a phantom or quintessence equation of state. To circumvent early-time superhorizon instabilities, the energy-momentum transfer should occur in opposing directions within the two regimes, resulting in distinct phenomenological outcomes. We study quintessence and phantom non-dynamical and dynamical models in light of two independent Cosmic Microwave Background (CMB) experiments  -- the Planck satellite and the Atacama Cosmology Telescope. We analyze CMB data both independently and in combination with Supernovae (SN) distance moduli measurements from the \textit{Pantheon-Plus} catalog and Baryon Acoustic Oscillations (BAO) from the SDSS-IV eBOSS survey. Our results update and extend the state-of-the-art analyses, significantly narrowing the parameter space allowed for these models and limiting their overall ability to reconcile cosmological tensions. Although considering different combinations of data leaves some freedom to increase $H_0$ towards the value measured by the SH0ES collaboration, our most constraining dataset (CMB+BAO+SN) indicates that fully reconciling the tension solely within the framework of IDE remains challenging.
\end{abstract}


\maketitle

\section{Introduction}
\label{sec:intro}

In spite of the successes accumulated in past decades, the standard $\Lambda$CDM model of cosmology seems to be struggling to provide an exhaustive description of the most recent observations. As experimental precision has improved, various anomalies and tensions between experiments have come to light~\cite{DiValentino:2020zio,DiValentino:2020vvd,Perivolaropoulos:2021jda,Abdalla:2022yfr}. Among them, one in particular seriously calls into question the validity of our best-working model of the Universe: the so-called Hubble tension~\cite{Verde:2019ivm,DiValentino:2021izs,Kamionkowski:2022pkx,Khalife:2023qbu}. 
 
The Hubble tension refers to a $\sim 5\sigma$ disagreement between the value of the Hubble constant as inferred by CMB data from the Planck collaboration~\cite{Planck:2018vyg} assuming a $\Lambda$CDM cosmology ($H_0=67.4\pm0.5$ km/s/Mpc) and the value of the same parameter as directly obtained by local distance ladder measurements from Type Ia supernovae from the SH0ES collaboration~\cite{Riess:2021jrx,Murakami:2023xuy} ($H_0=73\pm1$ km/s/Mpc). Barring any possible systematic origin of the tension,\footnote{Although it cannot be ruled out entirely, this possibility is becoming increasingly unlikely, given the extensive analysis performed by the SH0ES collaboration~\cite{Riess:2021jrx} and the common pattern observed in the distribution of other local and early time independent measurements of $H_0$.} an exciting possibility to consider is that this problem could represent an indication of new physics beyond the standard cosmological paradigm.

In this regard, we note that a somewhat surprising outcome of modern cosmology is that merely 5\% of the total energy density of the Universe comprises baryonic matter, whose physical properties are fairly well understood in the framework of the Standard Model of particle physics (SM). In stark contrast, the remaining 95\% is attributed to enigmatic entities -- Dark Energy (DE) and Dark Matter (DM) -- whose origins remain an enigma for modern cosmology and high-energy physics. Within the $\Lambda$CDM model, we opt for a quite minimal parameterization, describing DM as a perfect fluid made of cold non-relativistic particles with low momenta that do not interact with the other SM particles except through gravity. In addition, we assume DE to be a cosmological constant ($\Lambda$) in the Einstein Equation. However, it seems natural to speculate that, given our limited understanding of DM and DE, a part of the aforementioned tensions could originate from an oversimplification of the theoretical parameterization adopted for the dark sector of the cosmological model. For this reason, several intriguing alternatives involving additional interactions or couplings in the dark sector have been considered.

A model among many that has gained some research attention is the so-called Interacting Dark Energy (IDE) cosmology~\cite{Valiviita:2008iv,Gavela:2009cy,Bamba:2012cp,Salvatelli:2014zta,Nunes_2014,Escudero:2015yka,Sola:2016jky,Wang:2016lxa,Kumar:2016zpg,Murgia:2016ccp,Pourtsidou:2016ico,SolaPeracaula:2017esw,DiValentino:2017iww,Kumar:2017dnp,Sola:2017znb,Gomez-Valent:2018nib,Yang:2018euj,Yang:2018ubt,Barros:2018efl,Martinelli:2019dau,Yang:2019uog,DiValentino:2019ffd,Pan:2019jqh,Kumar:2019wfs,Yang:2019uzo,Pan:2019gop,DiValentino:2019jae,DiValentino:2020leo,Yao:2020pji,Lucca:2020zjb,DiValentino:2020kpf,Gomez-Valent:2020mqn,Yang:2020uga,Yao:2020hkw,Pan:2020bur,Pan:2020mst,DiValentino:2020vnx,Hogg:2020rdp,SolaPeracaula:2021gxi,Lucca:2021dxo,Kumar:2021eev,Yang:2021hxg,Gao:2021xnk,Yang:2021oxc,Lucca:2021eqy,Halder:2021jiv,Gariazzo:2021qtg,Nunes:2021zzi,Paliathanasis:2021egx,Bonilla:2021dql,Kaneta:2022kjj,Chatzidakis:2022mpf,Yang:2022csz,Nunes:2022bhn,Goh:2022gxo,Gomez-Valent:2022bku,
vanderWesthuizen:2023hcl,Zhai:2023yny,Bernui:2023byc,deCruzPerez:2023wzd,Escamilla:2023shf,Hoerning:2023hks,Forconi:2023hsj}. At its core, the model postulates a non-gravitational interaction between DM and DE, allowing an exchange of energy-momentum between the two, see also Refs.~\cite{Wang:2016lxa,Wang:2024vmw} for reviews. 

It is worth noting that, from a purely theoretical point of view, there is no fundamental symmetry in nature for which non-gravitational DM-DE interactions are forbidden. Cosmological models featuring an interacting dark sector (in part motivated by the idea of coupled quintessence~\cite{Wetterich:1994bg,Amendola:1999dr,Amendola:1999er,
Mangano:2002gg,Zhang:2005rg,Saridakis:2010mf,Barros:2018efl,
DAmico:2018mnx,Liu:2019ygl} ) have been largely explored and discussed in the literature, see, e.g.,~\cite{Farrar:2003uw,Barrow:2006hia,Amendola:2006dg,He:2008tn,Valiviita:2008iv,
Gavela:2009cy,CalderaCabral:2009ja,Majerotto:2009np,Abdalla:2009mt,Honorez:2010rr,
Clemson:2011an,Pan:2012ki,Salvatelli:2013wra,Nunes:2014qoa,
Faraoni:2014vra,Pan:2014afa,vandeBruck:2015ida,Tamanini:2015iia,Li:2015vla,Murgia:2016ccp,Nunes:2016dlj,Yang:2016evp,Pan:2016ngu,Sharov:2017iue,An:2017kqu,Santos:2017bqm,
Mifsud:2017fsy,Kumar:2017bpv,Guo:2017deu,Pan:2017ent,An:2017crg,Costa:2018aoy,
Wang:2018azy,vonMarttens:2018iav,Yang:2018qec,Martinelli:2019dau,Li:2019loh,
Yang:2019vni,Bachega:2019fki,Yang:2019uzo,Li:2019ajo,Mukhopadhyay:2019jla,Carneiro:2019rly,Kase:2019veo,Yamanaka:2019aeq,Yamanaka:2019yek,Mifsud:2019fut,vandeBruck:2020fjo,vandeBruck:2022xbk,Teixeira:2022sjr,Teixeira:2023zjt}. Furthermore, many have speculated that a coupled dark sector could help address the coincidence (or "why now?") problem~\cite{Hu:2006ar,Sadjadi:2006qp,delCampo:2008jx,Dutta:2017kch,Dutta:2017fjw}. 

On the other hand, from a more observational standpoint, IDE cosmology has emerged as a possible solution to cosmological tensions~\cite{Salvatelli:2014zta,Kumar:2016zpg,Costa:2016tpb,Xia:2016vnp,Kumar:2017dnp,Yang:2017ccc,Feng:2017usu,Yang:2018ubt,Zhang:2018glx,Yang:2018xlt,Yang:2018uae,Li:2018ydj,Kumar:2019wfs,Pan:2019jqh,DiValentino:2017iww,Yang:2017ccc,Feng:2017usu,Yang:2018ubt,Yang:2018xlt,Yang:2018uae,Li:2018ydj,Kumar:2019wfs,Pan:2019jqh,Yang:2019uzo,Pan:2019gop,Benetti:2019lxu,DiValentino:2019ffd,DiValentino:2019jae,Liu:2022hpz,Zhao:2022ycr,Zhai:2023yny,Pan:2023mie,Benisty:2024lmj,Giare:2024smz,Sabogal:2024yha}. Allowing an exchange of energy-momentum from DM to DE, can increase the value of $H_0$ inferred from CMB observations, possibly restoring the agreement with the direct measurement provided by the SH0ES collaboration. In addition, as recently shown by some of us in Ref.~\cite{Zhai:2023yny}, IDE seems to be supported by different independent CMB experiments, leading to an overall consistency of view concerning the amount of energy-momentum transferred in the dark sector. Having said that, whether or not this model could provide a successful solution for the Hubble trouble is still a matter of debate. The model suffers from the typical problem faced by any late-time solution (i.e., solutions that are aimed to solve the Hubble tension by introducing new physics post-recombination). Namely, Baryon Acoustic Oscillation (BAO) data and distance moduli measurements for Supernovae (SN) are very constraining on local distances, leaving us with little freedom to introduce any deviation from a basic $\Lambda$CDM late-time cosmology~\cite{Krishnan:2021dyb,Keeley:2022ojz}. As a result, when considering low-redshift ($z$) probes, the ability of IDE to increase the present-day expansion rate of the Universe is strongly reduced, if not completely lost.\footnote{It is worth mentioning some caveats surrounding the use of BAO data. Firstly, volumetric BAO data might retain a residual model dependence from the template used in the analysis pipeline. In addition, as pointed out by some of us in Ref.~\cite{Bernui:2023byc}, volumetric BAO data produce somewhat conflicting constraints on IDE compared to transverse 2D-BAO measurements, providing another element of concern.}

To better understand the extent to which IDE (and their relatives) can provide an actual solution to the $H_0$-tension, in this work, we focus more closely on the role played by the DE equation of state (EoS) in IDE cosmology. First and foremost, we note that the nature of the DE EoS acquires primary importance in the model. We will be more precise on this in the next sections (specifically in \autoref{sec:theory} and \autoref{sec:method}), but we anticipate that to avoid early-time superhorizon instabilities with cosmological perturbations~\cite{He:2008si,Jackson_2009,Gavela:2010tm}, the EoS must be theoretically confined to either the quintessence or phantom regime, depending on the direction in which energy-momentum is transferred between DM and DE. In the scenario we will consider in this work, a quintessential DE EoS ($w_0 > -1$) implies an energy-momentum flow from DM to DE. Conversely, a phantom DE EoS ($w_0<-1$) implies a transfer of energy-momentum from DE to DM.

An implicit assumption underlying a large portion of the results mentioned thus far is fixing the DE EoS to a very tiny quintessence value, $w_0 \simeq -0.999 \simeq -1$, essentially resembling a cosmological constant. Nevertheless, a few scattered analyses (largely conducted by some of us), have already considered the possibility of leaving the EoS $w_0$ a free parameter of the model~\cite{DiValentino:2019jae,Anchordoqui:2021gji,Yang:2021hxg,Yang:2022csz}. We refer to scenarios featuring a non-dynamical $w_0\ne -1$ EoS as $w_0$IDE. For these models, many important aspects remain unclear, and important questions are pending. For example, in Ref.~\cite{DiValentino:2019jae}, non-dynamical models were examined in the context of Planck-2018 data along with Baryon Acoustic Oscillations (BAO) and (SH0ES calibrated) Supernovae (SN) data. This analysis revealed a significant preference for quintessence IDE, establishing the model as a highly promising solution to the Hubble tension. However, as far as we know, the state-of-the-art constraints on $w_0$IDE remain largely unchanged since 2019, and such preferences have not been tested subsequently with CMB data other than Planck or against the latest low-redshift probes. Therefore, as a first step, we undertake a comprehensive reassessment of the constraints on the non-dynamical $w_0$IDE scenario, updating the constraints on the $w_0$IDE model (for both the quintessence and phantom regime), and incorporating the latest BAO, SN, and CMB data in the analysis. In this regard, we aim to clarify the following important aspects:

\begin{itemize}[leftmargin=*]
\item We place special emphasis on the constraints arising from (updated) local distance measurements, in the form of SN and BAO measurements to examine if any leeway remains to address cosmological tensions through DM-DE energy-momentum transfer (in either the quintessence or phantom regimes). In this context, interesting aspects to clarify are whether the latest BAO and SN data independently validate or dismiss the $w_0$IDE scenario as a viable solution to the Hubble tension and shed light on the role of the SH0ES calibration for SN. 
	
\item As we already mentioned, in Ref~\cite{Zhai:2023yny}, some of us pointed out that different CMB experiments share a consistent view on IDE when $w_0 \simeq -1$ is fixed. Here we extend the analysis of small-scale CMB measurements released by the Atacama Cosmology Telescope (ACT) to the case where $w_0$ is let free to vary. In this regard, a particularly relevant aspect to clarify is whether the well-documented Planck preference for a phantom DE EoS~\cite{Planck:2018vyg,Escamilla:2023oce} (not confirmed by ACT~\cite{ACT:2020gnv,DiValentino:2022oon,Giare:2023xoc}) could play any effect on the amount of energy-momentum transfer supported by data and, more broadly, if independent CMB experiments validate or dismiss the $w_0$IDE scenario as a viable solution to the Hubble tension.

\end{itemize}

We then move to consider more exotic models where the DE EoS $w(z)$ is dynamical and changes with cosmic expansion. We refer to this scenario as $w(z)$IDE. Notice that the interaction between DM and DE via dynamical DE EoS has been investigated in various other contexts, see e.g., Refs.~\cite{Nunes_2014, vandeBruck:2015ida,Yang:2022csz}. 
Here, we describe the dynamical evolution of $w(z)$ by adopting a simple Chevallier-Polarski-Linder (CPL) parametrization~\cite{CHEVALLIER_2001,Linder_2003} and provide a comprehensive overview of the most recent observational constraints on the dynamical $w(z)$IDE cosmology. We are fueled by the following motivations:

\begin{itemize}[leftmargin=*]
	
\item To the best of our knowledge, a proper updated analysis aimed to understand whether the dynamical $w(z)$IDE model could represent a solution to the Hubble tension (either in the quintessence regime or in the phantom regime) is missing. Therefore, we believe it is intrinsically interesting to understand how a possible dynamic behavior of the DE EoS could impact the constraints derived for the non-dynamical case.

\item From a more practical point of view, the questions we seek to answer are not different from what we already pointed out: interesting aspects to clarify are whether the latest BAO and SN data independently validate or dismiss the $w(z)$IDE scenario as a viable solution to the $H_0$ tension. Additionally, we want to assess whether ACT and Planck always share a consistent view on $w(z)$IDE.
	
\end{itemize}

The paper is organized as follows. In \autoref{sec:theory}, we introduce the theoretical framework that underpins our study. In \autoref{sec:method}, we outline the methodology and the updated datasets used to establish observational constraints within the models considered in this work. Moving further, \autoref{sec:Results_non-DDE} and \autoref{sec:Results_DDE} delve into our primary findings for the non-dynamical and dynamical cases, respectively. Finally, in \autoref{sec:conclusions}, we summarize our conclusions and offer insights into future perspectives. As usual, a subindex zero attached to any quantity means that it must be evaluated at present time.

\section{Interacting Dark Energy Cosmology}
\label{sec:theory}

In this section, we review in a nutshell the well-established aspects of background evolution and linear perturbations that govern the coupling between two dark fluids.  We consider that the gravitational sector of the universe is described by the Einstein's General Relativity and a flat Friedmann-Lema\^{i}tre-Robertson-Walker (FLRW) line element in the synchronous gauge $ds^2 = a^2 (\eta) \left[- d \eta^2 + (\delta_{ij} + h_{ij}) dx^i dx^j\right]$, where $a(\eta)$ is the scale factor as a function of the conformal time $\eta$; $\delta_{ij}$, $h_{ij}$ are respectively the unperturbed and perturbed metric tensors. The conservation equations of DM and DE in the presence of a dark interaction, characterized by a coupling function $Q(t)$, can be expressed as follows:

\begin{eqnarray}
&& \nabla_{\nu} T^{\mu \nu}_{\rm DM} =  \frac{Q  u^{\mu}}{a (\eta)},\\
&& \nabla_{\nu} T^{\mu \nu}_{\rm DE} =  - \frac{Q u^{\mu}}{a (\eta)},
\end{eqnarray}
where $T^{\mu \nu}_{\rm DM}$ and $T^{\mu \nu}_{\rm DE}$ are respectively the energy-momentum tensor for DM and DE; $u^{\mu}$ is the four-velocity vector of DM which in the synchronous gauge can be defined as $u^{\mu} = a (\eta) (-1, u^{i})$ where $\mu =0, 1, 2, 3$, $i = 1, 2, 3$ and $u^i$ is the proper velocity of the DM fluid. Now, to understand the evolution of the dark fluids, at the background and perturbation levels, one needs to prescribe the nature of the dark fluids and also the interaction function.
We assume that DM is pressureless, with $\rho_c$ denoting its energy density, while DE enjoys a (dynamical or non-dynamical) barotropic equation of state, represented by $w_x = p_x/\rho_x$, where $p_x$ and $\rho_x$ are respectively the pressure and energy density of the DE fluid. By examining the sign of $Q(t)$, we can determine the direction of energy transfer between the dark components. When $Q > 0$, energy transfers from DE to cold dark matter (CDM), while $Q < 0$ indicates the reverse situation, signifying energy transfer from CDM to DE.

Given the complexity of describing both dark species, at this stage, to proceed, an exact phenomenological approach that quantifies the coupling must be assumed. Several proposals in this regard have been put forward in the literature in recent times. In this article, we consider a widely studied model of the interaction function, which has received considerable attention in recent years. The interaction rate that we employ is as follows:

\begin{eqnarray}\label{model-Q}
    Q = \xi \mathcal{H} \rho_x,
    \label{eq:xi}
\end{eqnarray}
where $\xi$ represents the dimensionless coupling constant which is independent of cosmic time. Here, $\mathcal{H}$ denotes the conformal Hubble parameter, following the standard definition as in General Relativity: $3 \mathcal{H}^2 = 8\pi G\; a^2 (\eta) \sum \rho_i$, where $\rho_i$ represents the energy density of the $i$-th fluid. In addition to  CDM and DE, we have also accounted for the presence of baryons, radiation, and neutrinos, including one massive and two massless species. Hence, consistent with the sign convention of $Q(t)$, $\xi > 0$ ($< 0$) denotes the transfer of energy from DE to CDM (from CDM to DE), respectively. \footnote{It is worth noting that Ref. \cite{Pan:2020zza} explicitly demonstrated how interaction rates, as considered in Eq. (\ref{model-Q}), can naturally arise from first principles when exploring well-motivated field theories for scenarios of IDE.}

Now, considering the linear perturbations and to prevent any potential unphysical scenarios related to the DE equation of state and $c_{s,x}^2$, we set the DE sound speed $c_{s,x}^2 = 1$. This allows us to express the evolution of density perturbations in terms of $\delta_{c,x}$ and velocity perturbations ($\theta_{c,x}$) as follows:

\small
\begin{eqnarray}
\delta'_x &=& -(1+w_x) \left ( \theta_x+\frac{h'}{2} \right )-\xi \left ( \frac{kv_T}{3}+\frac{h'}{6} \right ) \nonumber \\
&&-3{\cal H}(1-w_x) \left [ \delta_x+\frac{{\cal H}\theta_x}{k^2} \left (3(1+w_x)+\xi \right ) \right ]\,, \label{eq:deltax}\\
\theta'_x &=& 2{\cal H}\theta_x+\frac{k^2}{1+w_x}\delta_x+2{\cal H}\frac{\xi}{1+w_x}\theta_x-\xi{\cal H}\frac{\theta_c}{1+w_x}, \label{eq:thetax}\\
\delta'_c &=& -\theta_c - \frac{1}{2}h' +\xi{\cal H}\frac{\rho_x}{\rho_c}(\delta_x-\delta_c)+\xi\frac{\rho_x}{\rho_c} \left ( \frac{kv_T}
{3}+\frac{h'}{6} \right ), \label{eq:deltac} \\
\theta'_c &=& -{\cal H}\theta_c, \label{eq:thetac}
\end{eqnarray}
\normalsize
where $h$ is the trace of the scalar metric perturbation $h_{ij}$; $k$ is the Fourier-space wave number; the prime attached to any quantity denotes its derivative with respect to the conformal time and $v_T$ refers to the center of mass velocity of the total fluid~\cite{Gavela:2010tm}.

Ensuring the stability of linear perturbations over time is crucial within the dynamical scenarios considered for the dark coupling in this study. As demonstrated in~\cite{Gavela:2009cy}, the parameter known as the "doom factor", denoted as $\mathrm{d} = Q/ (3 \mathcal{H} (1+w_x))$, plays a pivotal role in determining the stability of the scalar modes. When $\mathrm{d} <0$, indicating stability, it implies that for $Q > 0$, the equation of state parameter $w_x$ must be less than $-1$. Conversely, for $Q < 0$, $w_x$ needs to be greater than $-1$ to maintain stability. In the following section, we will delve into a detailed description of the parameter space that ensures an absence of instabilities.

\section{Methodology and data}
\label{sec:method}

\subsection{Markov Chain Monte Carlo Analysis}
\label{sec:MCMC}

We perform Markov Chain Monte Carlo (MCMC) analyses using the publicly available sampler \texttt{COBAYA}~\cite{Torrado:2020dgo}. The code explores the posterior distributions of a given parameter space using the MCMC sampler developed for \texttt{CosmoMC}~\cite{Lewis:2002ah} and tailored for parameter spaces with speed hierarchy, implementing the "fast dragging" procedure detailed in Ref.~\cite{Neal:2005}. We compute the theoretical model by means of the Cosmic Linear Anisotropy Solving System code, \texttt{CLASS}~\cite{Blas:2011rf}, modified to introduce the possibility of interactions between dark energy and dark matter. 

Our sampling parameters are the usual six $\Lambda$CDM parameters, namely the baryon $\omega_{\rm b} \doteq \Omega_{\rm b}h^2$ and cold dark matter $\omega_{\rm c} \doteq \Omega_{\rm c}h^2$ energy densities, the angular size of the horizon at the last scattering surface $\theta_{\rm{MC}}$, the optical depth $\tau$, the amplitude of primordial scalar perturbation $\log(10^{10}A_{\rm s})$, and the scalar spectral index $n_s$. In addition, we consider the coupling parameter $\xi$ defined in Eq.~(\ref{eq:xi}) and the set of parameters describing the DE EoS. In this regard, we distinguish two different cases:

\begin{table}[t]
	\begin{center}
		\renewcommand{\arraystretch}{1.5}
		\begin{tabular}{c@{\hspace{0.3 cm}}@{\hspace{0.3 cm}} c @{\hspace{0.3cm}} c }
			\hline
			\textbf{Parameter}                       & \textbf{Phantom}  & \textbf{Quintessence}\\
			\hline\hline
			$\Omega_{\rm b} h^2$                     & $[0.005\,,\,0.1]$ &$[0.005\,,\,0.1]$\\
			$\Omega_{\rm c} h^2$         & $[0.01\,,\,0.99]$ &$[0.01\,,\,0.99]$\\
			$100\,\theta_{\rm {MC}}$                 & $[0.5\,,\,10]$ &$[0.5\,,\,10]$\\
			$\tau$                                   & $[0.01\,,\,0.8]$ & $[0.01\,,\,0.8]$\\
			$\log(10^{10}A_{\rm S})$                 & $[1.61\,,\,3.91]$ & $[1.61\,,\,3.91]$\\
			$n_{s}$                                  & $[0.8\,,\, 1.2]$ & $[0.8\,,\, 1.2]$\\
            $w_0$                                    & $[-3\,,\, -1)$ & $(-1\,,\, 1]$\\
            $w_a$                                    & $[-3\,,\, 2]$ & $[-3\,,\, 2]$\\
			$\xi$                                    & $[0\,,\,1]$ & $[-1\,,\,0]$\\
			
			\hline\hline
		\end{tabular}
		\caption{List of the uniform parameter priors for the phantom and quintessence regimes. When considering the ACT CMB data, we assume a Gaussian prior $\tau= 0.065 \pm 0.015$ with a width much smaller than the uniform prior reported in this table.}
		\label{tab.Priors}
	\end{center}
	
\end{table}

\begin{itemize}[leftmargin=*]

\item \textbf{Non-Dynamical DE EoS --} In this case $w_0$ becomes an additional free parameter that we vary in two different regimes: the quintessence regime where $w_0 > -1$ and the phantom regime where $w_0 < -1$. To avoid instabilities in primordial perturbations, the coupling parameter $\xi$ can vary within the following priors: $\xi < 0$ when $w_0 > -1$ and $\xi > 0$ when $w_0 < -1$.\footnote{Note that in the quintessence case, the upper prior limit is set at $w_0 < 1$. This is a common choice widely used in the literature, and can be partly justified by the fact that quintessence models are typically based on scalar field realizations. Within the minimal theoretical framework, the EoS cannot exceed $w_0 = 1$, providing a physical rationale for this upper limit, while also allowing sufficient margin for the necessary condition to achieve acceleration, $w_0 < -\frac{1}{3}$.} Therefore the behavior of the DE EoS fixes the direction of energy-momentum transfer between DM and DE. In summary:\\
\\
\textit{i)} $w_0>-1 \Leftrightarrow \xi<0 \Leftrightarrow \text{DM to DE}$,\\
\\
\textit{ii)} $w_0<-1 \Leftrightarrow \xi>0 \Leftrightarrow \text{DE to DM}$.\\
\\

\item \textbf{Dynamical DE EoS -- } In this case, we adopt a CPL parameterization:
\begin{equation}
w(a) = w_0 + w_a (1-a),
\label{eq:CPL}
\end{equation}
where $w_0$ represents the present value $w(a=1)$, and $w_a$ is another free parameter such that $\frac{dw}{d\ln(1+z)}\bigg|_{z=1}=\frac{w_a}{2}$. As usual, we distinguish two different regimes:\\
\\
\textit{i)} The quintessence regime where $w(z) > -1$ for any $z$. In the MCMC analysis, we sample over these two parameters, ensuring that for every randomly sampled pair of values $w_0$-$w_a$, the condition $w(z) > -1$ holds true at any $z$.\footnote{Previous studies typically imposed priors on the CPL parameterization's free parameters to force the model into the phantom or quintessence regime; see, e.g., Ref.~\cite{Yang:2022csz}, where some of us considered the CPL parameterization in similar yet distinct IDE cosmologies. In contrast, our approach ensures that $w(z)$ can always lie in the quintessence ($w(z) > -1$) or phantom ($w(z) < -1$) regime at any $z$ without assuming priors but by dynamically checking $w(z)$ during the MCMC. Although this may seem like a technical detail within the CPL parameterization (where both approaches yield similar results), it becomes crucial when studying dynamical dark energy models beyond CPL (which is not done in this study). In such cases, conditions on $w(z)$ cannot always be easily mapped into priors on parameters, and our method allows for proper sampling of the parameter space, which has not been explored in the literature.} If this condition is not met, the point is rejected. The test is performed dynamically during the MCMC run, without assuming any restrictive prior on the parameter $w_a$ controlling the dynamical evolution of $w(z)$. In contrast, we assume a prior $w_0>-1$ that automatically follows from requiring $w(z) > -1$ at $z=0$. Our methodology ensures proper sampling of the parameter space and convergence of the chains.\\
\\
\textit{ii)} The phantom regime where $w(z) < -1$ for any $z$. We ensure that for every sampled pair of values $w_0$-$w_a$ during the MCMC run, the condition $w(z) < -1$ is satisfied at any $z$. If this condition is not met, the point is rejected. Also in this case the test is performed dynamically without assuming any restrictive prior on $w_a$ and imposing a prior $w_0<-1$ that automatically follows from requiring $w(z) < -1$ at $z=0$.\\
\\
As for the coupling parameter $\xi$, to avoid instabilities in primordial perturbations, in the quintessence regime we need $\xi < 0$, while in the phantom case $\xi > 0$. Therefore, also in the dynamical case, the DE EoS regime determines the direction of the energy and momentum flow between DM and DE. In summary:
\\
\\
\textit{i)} $w(z)>-1 \, \forall \, z \Leftrightarrow \xi<0 \Leftrightarrow \text{DM to DE}$\\
\\
\textit{ii)} $w(z)<-1 \, \forall \, z \Leftrightarrow \xi>0 \Leftrightarrow \text{DE to DM}$
\\
\end{itemize}

A summary of the uniform prior distributions adopted for all the cosmological parameters considered in the analysis is given in \autoref{tab.Priors} (except for the optical depth at reionization $\tau$ for which we adopt a prior distribution that depends on the specific CMB dataset, as discussed below). We test the convergence of the chains obtained using this approach by means of the Gelman-Rubin criterion. We establish a threshold for chain convergence of $R-1 \lesssim 0.02$.

\subsection{Cosmological Data}
\label{sec:Data}

Our reference datasets for both the dynamical and non-dynamical IDE scenarios are the following:
\begin{itemize}[leftmargin=*]
    \item The Planck 2018 temperature and polarization (TT TE EE) likelihood, which also includes low multipole data ($\ell < 30$)~\cite{Planck:2019nip,Planck:2018vyg,Planck:2018nkj} and the Planck 2018 lensing likelihood~\cite{Planck:2018lbu}, constructed from measurements of the power spectrum of the lensing potential. We refer to this dataset as \textit{\textbf{P18}}.
    
    \item Atacama Cosmology Telescope temperature and polarization anisotrpy DR4 likelihood in combination with  the gravitational lensing DR6 likelihood covering 9400 deg$^2$ reconstructed from CMB measurements made by the Atacama Cosmology Telescope from 2017 to 2021~\cite{ACT:2023kun,ACT:2023dou}. In our analysis for the lensing spectrum we include only the conservative range of lensing multipoles $40 < \ell < 763$.  We consider a Gaussian prior on $\tau = 0.065 \pm 0.015$, as done in~\cite{ACT:2020gnv}. We refer to this dataset as \textit{\textbf{ACT}}.
    
    \item Baryon Acoustic Oscillation data from the finalized SDSS-IV eBOSS survey. These data encompass both isotropic and anisotropic distance and expansion rate measurements, as outlined in Table 3 of Reference~\cite{eBOSS:2020yzd}. We refer to this dataset as \textbf{\textit{BAO}}.

    \item Distance modulus measurements of Type Ia supernovae obtained from the \textit{Pantheon-Plus} sample~\cite{Brout:2022vxf}. This dataset comprises 1701 light curves representing 1550 unique Type Ia supernovae, spanning a redshift range from 0.001 to 2.26. In all our analyses, we consider two distinct possibilities. On the one hand, we consider the uncalibrated \textit{Pantheon-Plus} SNe Ia sample that we will henceforth refer to as \textbf{\textit{SN}}. On the other hand, we will consider the SH0ES Cepheid host distances to calibrate the SNe Ia sample~\cite{Riess:2021jrx}. We refer to the SH0ES-calibrated SN dataset as \textbf{\textit{SN+SH0ES}}.    
\end{itemize}

\section{Results for non-Dynamical EoS}
\label{sec:Results_non-DDE}

In this section, we present the results obtained considering a non-Dynamical DE EoS. We divide the section into two different subsections. In \autoref{sec:qIDE}, we focus on quintessence models characterized by $w_0 > -1$ and a negative DM-DE coupling $\xi < 0$. These models feature a flow of energy-momentum from the DM sector to the DE sector of the theory. Conversely, in \autoref{sec:pIDE}, we study phantom models with $w_0 < -1$ and $\xi > 0$. In this case the energy momentum is transferred from DE to DM.

\subsection{Quintessence EoS}
\label{sec:qIDE}

The results obtained considering a quintessence DE EoS are provided in \autoref{tab.results.nondyanmical.quintessence.P18} and \autoref{tab.results.nondyanmical.quintessence.ACT}. In particular, \autoref{tab.results.nondyanmical.quintessence.P18} focuses on P18 temperature polarization and lensing data on their own and in different combinations with BAO and SN measurements while in \autoref{tab.results.nondyanmical.quintessence.ACT} we present the results obtained considering the small-scale CMB temperature polarization and lensing data released by ACT (DR4 and DR6), always on their own and in combinations with BAO and SN. In what follows we summarize the most interesting findings.
	
\begin{table*}[htpb!]
\begin{center}
\renewcommand{\arraystretch}{1.5}
\resizebox{0.9 \textwidth}{!}{
\begin{tabular}{l c c c c c c c c c c c c c c c }
\hline
\textbf{Parameter} & \textbf{ P18 } & \textbf{ P18+SN } & \textbf{ P18+SN+SH0ES } & \textbf{ P18+BAO } & \textbf{ P18+BAO+SN } \\ 
\hline\hline

$ \Omega_\mathrm{b} h^2  $ & $  0.02238\pm 0.00015 $ & $  0.02235\pm 0.00014 $ & $0.02252\pm 0.00014$ & $  0.02243\pm 0.00014 $ & $  0.02243\pm 0.00014 $ \\

$ \Omega_\mathrm{c} h^2  $ & $  0.071\pm 0.032 $ & $  0.098^{+0.022}_{-0.015} $ & $0.072^{+0.026}_{-0.012}$ & $  0.1117\pm 0.0039 $ & $  0.1122^{+0.0044}_{-0.0040} $ \\ 

$ 100\theta_\mathrm{s}  $ & $  1.04191\pm 0.00029 $ & $  1.04185\pm 0.00029 $ & $1.04208\pm 0.00029$ & $  1.04195\pm 0.00029 $ & $  1.04195\pm 0.00029 $ \\ 

$ \tau_\mathrm{reio}  $ & $  0.0560\pm 0.0066 $ & $  0.0544\pm 0.0067 $ & $5.7872^{+0.0072}_{-0.0080}$ & $  0.0580\pm 0.0077 $ & $  0.0580\pm 0.0073 $ \\ 

$ n_\mathrm{s}  $ & $  0.9658\pm 0.0044 $ & $  0.9644\pm 0.0040 $ & $0.9690\pm0.0040$ & $  0.9671\pm 0.0038 $ & $  0.9669\pm 0.0036 $ \\ 

$ \log(10^{10} A_\mathrm{s})  $ & $  3.047\pm 0.013 $ & $  3.045\pm 0.013 $ & $3.049\pm 0.015$ & $  3.050\pm 0.015 $ & $  3.050\pm 0.014 $ \\

$ \xi  $ & $ -0.39^{+0.36}_{-0.13}\, (> -0.759 ) $ & $ > -0.475 $ & $-0.37^{+0.18}_{-0.11}\, (-0.37^{+0.26}_{-0.31})$ & $  -0.067^{+0.045}_{-0.029}\, (> -0.126 ) $ & $  -0.063^{+0.048}_{-0.026}\, (> -0.123 )$ \\ 

$ w_0  $ & $ < -0.787 $ & $ < -0.815 $ & $<-0.843$ & $ < -0.920 $ & $ < -0.915 $ \\ 

$ H_0  $ [km/s/Mpc] & $  68.8\pm 3.2 $ & $  66.56\pm 0.76 $ & $69.75\pm 0.67$ & $  67.40\pm 0.66 $ & $  67.13\pm 0.57 $ \\ 

$ \Omega_\mathrm{m}  $ & $  0.204\pm 0.083 $ & $  0.274^{+0.050}_{-0.035} $ & $0.197^{+0.054}_{-0.029}$ & $  0.297\pm 0.011 $ & $  0.300\pm 0.010 $ \\

$ \sigma_8  $ & $  1.50^{+0.51}_{-1.1} $ & $  0.98^{+0.17}_{-0.26} $ & $1.299^{+0.043}_{-0.385}$ & $  0.848\pm 0.025 $ & $  0.843^{+0.024}_{-0.027} $ \\ 

$ r_\mathrm{drag}  $ & $  147.16\pm 0.28 $ & $  147.10\pm 0.25 $ & $147.35\pm 0.25$ & $  147.31\pm 0.23 $ & $  147.30\pm 0.23 $ \\ 

\hline \hline
\end{tabular} }
\end{center}
\caption{Constraints at 68\% (95\%) CL and upper limits at 95\% CL on the parameters of the $w_0$IDE model with $w_0>-1$. The results are obtained by different combinations of P18, BAO and SN (with and without the SH0ES calibration) measurements.}
\label{tab.results.nondyanmical.quintessence.P18}
\end{table*}

\begin{table*}[htpb!]
\begin{center}
\renewcommand{\arraystretch}{1.5}
\resizebox{0.9\textwidth}{!}{
\begin{tabular}{l c c c c c c c c c c c c c c c }
\hline
\textbf{Parameter} & \textbf{ ACT } & \textbf{ ACT+SN } & \textbf{ ACT+SN+SH0ES } & \textbf{ ACT+BAO } & \textbf{ ACT+BAO+SN } \\ 
\hline\hline

$ \Omega_\mathrm{b} h^2  $ & $0.02161\pm0.00030$ & $  0.02162\pm 0.00031 $ & $0.02193\pm0.00030$ & $  0.02166\pm 0.00029 $ & $  0.02165\pm 0.00030 $ \\ 

$ \Omega_\mathrm{c} h^2  $ & $<0.112$ & $  0.062^{+0.028}_{-0.047} $ & $<0.096$ & $  0.1126^{+0.0043}_{-0.0037} $ & $  0.1127^{+0.0045}_{-0.0038} $ \\ 

$ 100\theta_\mathrm{s}  $ & $1.04324\pm0.00067$ & $  1.04320\pm 0.00067 $ & $1.04378\pm 0.00065$ & $  1.04334\pm 0.00063 $ & $  1.04330\pm 0.00065 $ \\ 

$ \tau_\mathrm{reio}  $ & $0.068\pm 0.014$ & $  0.068\pm 0.014 $ & $8.264\pm 0.014$ & $  0.077\pm 0.012 $ & $  0.077\pm 0.012 $ \\

$ n_\mathrm{s}  $ & $0.996\pm 0.012$ & $  0.995\pm 0.012 $ & $0.996\pm 0.012$ & $  0.996\pm 0.012 $ & $  0.996\pm 0.012 $ \\ 

$ \log(10^{10} A_\mathrm{s})  $ & $3.068\pm 0.026$ & $  3.070\pm 0.026 $ & $3.096\pm0.025$ & $  3.084\pm 0.022 $ & $  3.084\pm 0.022 $ \\ 

$ \xi  $ & $-0.50^{+0.15}_{-0.30}\, (>-0.80)$ & $ -0.45\pm 0.23\, (> -0.786 ) $ & $-0.49^{+0.13}_{-0.25}\, (-0.49^{+0.34}_{-0.29})$ & $ > -0.107 $ & $ > -0.110 $ \\ 

$ w_0  $ & $<-0.56$ & $ < -0.699 $ & $-0.869^{+0.077}_{-0.059}\,(<-0.775)$ & $ < -0.888 $ & $ < -0.898 $ \\ 

$ H_0  $ [km/s/Mpc] & $66.9^{+5.8}_{-3.5}$ & $  66.32\pm 0.90 $ & $69.96\pm 0.70$ & $  66.85^{+0.93}_{-0.84} $ & $  66.72\pm 0.61 $ \\ 

$ \Omega_\mathrm{m}  $ & $0.181^{+0.050}_{-0.130}$ & $  0.192\pm 0.075 $ & $0.146^{+0.044}_{-0.086}$ & $  0.302\pm 0.011 $ & $  0.303\pm 0.010 $ \\ 

$ \sigma_8  $ & $1.91^{+0.25}_{-1.15}$ & $  1.66^{+0.58}_{-1.2} $ & $1.95^{+0.26}_{-1.06}$ & $  0.856\pm 0.022 $ & $  0.854^{+0.022}_{-0.025} $ \\

$ r_\mathrm{drag}  $ & $148.03^{+0.71}_{-0.64}$ & $  148.06\pm 0.62 $ & $148.66\pm 0.60$ & $  148.43\pm 0.49 $ & $  148.43\pm 0.51 $ \\ 

\hline \hline
\end{tabular} }
\end{center}
\caption{Constraints at 68\% (95\%) CL and upper limits at 95\% CL on the parameters of the $w_0$IDE model with $w_0>-1$. The results are obtained by different combinations of ACT, BAO and SN (with and without the SH0ES calibration) measurements.}
\label{tab.results.nondyanmical.quintessence.ACT}
\end{table*}

\begin{figure*}[hpt]
    \centering
    \includegraphics[width=0.7\linewidth]{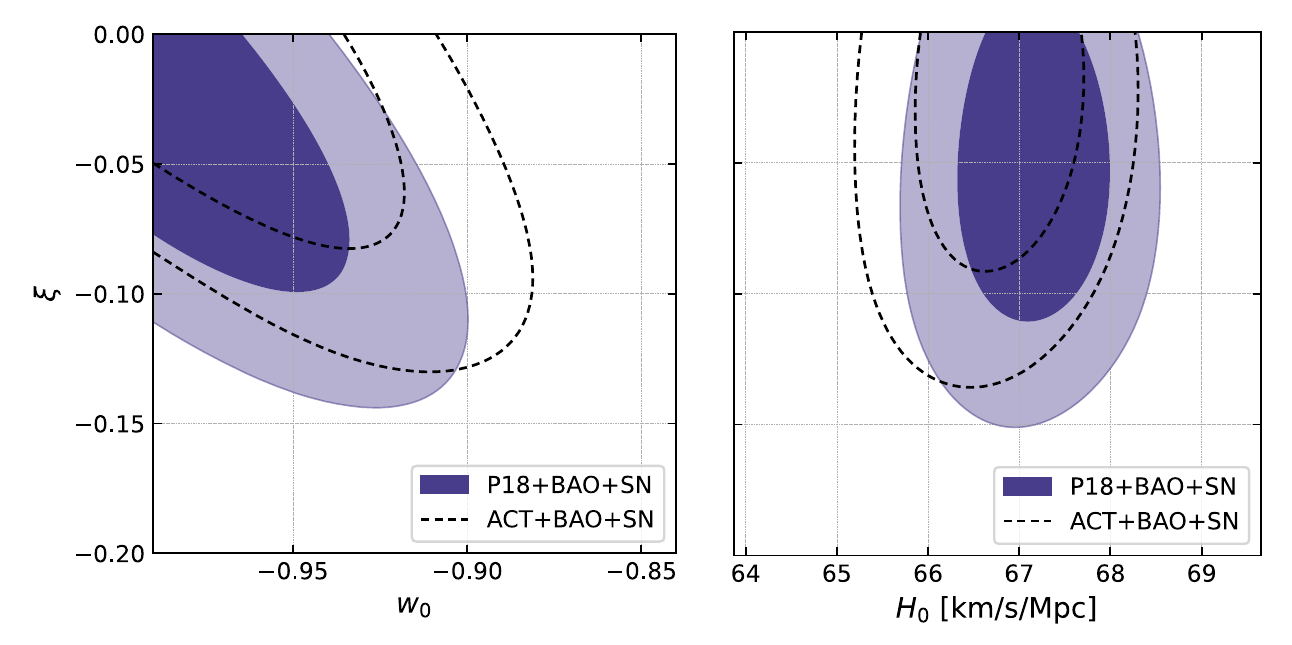}
    \caption{Joint marginalized contours at 68\% and 95\% CL illustrating the correlation between the coupling parameter $\xi$, the non-dynamical quintessence DE EoS $w_0$, and the Hubble parameter $H_0$ for P18+BAO+SN and ACT+BAO+SN.
}
    \label{fig:1}
\end{figure*}

\subsubsection{CMB-only}

Considering only CMB temperature, polarization, and lensing spectra, we have limited power to simultaneously constrain the DE EoS $w_0$ and the coupling parameter $\xi$. Concerning the former, from P18, we obtain an upper limit of $w_0 < -0.787$ at 95\% CL. Replacing P18 with ACT data further relaxes this upper limit to $w_0 < -0.56$. As for the coupling $\xi$, in both experiments we observe a 68\% CL preference for non-vanishing energy-momentum flow ($\xi = -0.39^{+0.36}_{-0.13}$ from P18 and $\xi = -0.39^{+0.15}_{-0.30}$ from ACT). However, this preference diminishes at the 95\% confidence level, and the analysis of both datasets indicates a broad lower bound of $\xi \gtrsim -0.8$, which lacks specificity and informative value. As expected, the main challenge arises from the so-called geometrical degeneracy among parameters. Essentially, different combinations of late-time cosmic parameters can be adjusted to keep the acoustic angular scale $\theta_s$ -- defined by the ratio of the comoving sound horizon at recombination to the comoving distance to last scattering -- constant as long as both quantities change proportionally. As a result, measurements based solely on CMB data, which can accurately determine this scale, cannot impose strong constraints on the (dynamical) IDE model by themselves unless late-time data are also incorporated to partially break this degeneracy.

When it comes to the present-day expansion rate, we obtain $H_0 = 68.8 \pm 3.2$ km/s/Mpc from Planck and $H_0 = 66.9^{+5.8}_{-3.5}$ km/s/Mpc from ACT. We can compare these results with those in Tab.I of Ref.~\cite{Zhai:2023yny} that were derived under the assumption $w_0\simeq -1$. This comparison reveals that allowing the EoS to freely vary in the cosmological model can have significant implications for the results. On the one hand, allowing an additional parameter to vary produces a significant increase in the uncertainties. This is largely expected when studying models featuring new physics at late times only in light of CMB data. The reason is that we face the well-known geometrical degeneracy problem, namely the fact that different combinations of parameters can be arranged to maintain the same CMB acoustic angular scale $\theta_s$. This degeneracy makes it challenging to disentangle their effects on the CMB spectra (unless perturbation-level effects are included). On the other hand, referring back to Ref.~\cite{Zhai:2023yny}, we notice that when $w_0$ is left free to vary in the quintessence regime, in both experiments, the fitting value of $H_0$ significantly shifts towards values closer to the one obtained within a $\Lambda$CDM model of cosmology. This shift is partly expected due to a simple argument: neglecting any interactions, it is a well-known fact that a quintessence EoS typically correlates with the present-day expansion rate of the universe in such a way that smaller values of $H_0$ are required to compensate for a (deep) quintessence $w_0$.

Summing up, when $w_0$ can vary in the quintessence regime, given the large uncertainties observed in both CMB experiments, it is difficult to derive definitive conclusions concerning the effective ability of the model to represent a valid solution to the Hubble constant tension. However, our analysis suggests from the onset that fixing $w_0$ to a value resembling the cosmological constant can represent an \textit{ansatz} for the model with non-negligible impact on the results.

\subsubsection{CMB and SN}

As a next step, to gain some constraining power, we incorporate SN data into our analysis. When dealing with SN, a decision needs to be made regarding whether to consider the uncalibrated dataset or introducing the SH0ES calibration for the absolute SN magnitude. On the one hand, a conservative approach would involve using the uncalibrated \textit{Pantheon-Plus} dataset and examining it alongside CMB measurements. On the other hand, if we take the results of the CMB-only analysis at face value, the Hubble tension is significantly reduced (mainly due to larger uncertainties). Therefore, using the SH0ES calibration is an alternative compelling decision in some specific cases.\footnote{In this regard, we would like to clarify some important aspects concerning the tension among datasets. As already pointed out in the text, when CMB data are analyzed alone, $H_0$ is poorly constrained. Due to the large uncertainties, the value of $H_0$ inferred from CMB data in this model is not necessarily in disagreement with SH0ES. This allows us to legitimately calibrate SN with SH0ES and analyze P18+SN+SH0ES to see whether the SH0ES calibration is supported by the model (i.e., to what extent we can increase $H_0$). On the other hand, the Hubble tension can be reframed as a tension among calibrators: SN calibrated with SH0ES and BAO calibrated with CMB, assuming standard early-time (i.e., pre-recombination) cosmology (as in the IDE model), are in strong tension (see Fig.~1 of Ref~\cite{Pogosian:2021mcs}). This means that combining CMB+BAO+SN+SH0ES would be inappropriate due to this tension. Indeed, we never consider such a combination of data. Conversely, when CMB+BAO+SN are analyzed together, the SH0ES calibration is never used. Therefore, we ensure that we never combine datasets that are in tension with each other while exploring all possible informative combinations of data.}

We start by considering uncalibrated SN in combination with CMB measurements. In this case, the 95\% CL constraints on the DE EoS become $w_0 < -0.815$ from P18+SN and $w_0 < -0.699$ for ACT+SN. The upper limits on the coupling $\xi$ are improved compared to the CMB-only case for P18+SN ($\xi > -0.475$), while they remain almost unchanged for ACT+SN ($\xi > -0.786$). However, the largest improvement in terms of constraining power is observed in the results on the Hubble parameter. The analysis of Planck+SN ($H_0 = 66.56 \pm 0.76$ km/s/Mpc) and ACT+SN ($H_0 = 66.32 \pm 0.90$ km/s/Mpc) agrees on values of $H_0$ that are in strong tension with respect to the local measurement provided by the SH0ES collaboration. As a result, taking the CMB+SN datasets at face value, the model would be unable to resolve the Hubble tension.

Having said that, it is worth considering that the situation looks very different when SN are calibrated with SH0ES.\footnote{We would like to cautiously remark that CMB~(P18/ACT)+SN and SN+SH0ES are in tension at more than $3\sigma$, highlighting the significant impact of assuming or not assuming a SH0ES calibration when dealing with SN measurements.} In this case, the constraints on the DE EoS become more restrictive on deviations away from the cosmological constant. We get $w_0<-0.843$ and $w_0<-0.775$ for P18+SN+SH0ES and ACT+SN+SH0ES, respectively. In addition, from both P18+SN+SH0ES ($\xi=-0.37^{+0.18}_{-0.11}$) and ACT+SN+SH0ES ($\xi=-0.49^{+0.13}_{-0.25}$), we observe a preference for a non-vanishing energy-momentum flow that is in remarkable agreement for the two experiments and persists at 95\% CL. This preference for a non-vanishing interaction produces higher values of the Hubble parameter ($H_0=69.75\pm0.67$ km/s/Mpc and $H_0=69.96\pm0.70$ km/s/Mpc for Planck+SN+SH0ES and ACT+SN+SH0ES, respectively). 

In conclusion, combining CMB observations with uncalibrated supernovae does not lead to an increase in the measured expansion rate of the universe. However, using a calibrated supernova dataset may result in a slight increase in the Hubble constant which is primarily driven by the SH0ES calibration. 

\subsubsection{CMB and BAO}

We now turn to the study of the effects of BAO data. As largely expected, BAOs are very constraining on deviations away from the cosmological constant. The combination of P18+BAO data produces upper limits $w_0<-0.920$, while from ACT+BAO we get $w_0<-0.888$, both at 95\% CL. Additionally, we strongly constrain the amount of energy-momentum that can be transferred from DM to DE. For P18+BAO and ACT+BAO, the constraints are improved all the way up to $\xi >-0.126$ and $\xi>-0.107$, always at 95\% CL. Easy to guess, the value of the Hubble parameter is now essentially the one predicted in the standard cosmological paradigm as we get $H_0=67.40\pm0.66$ km/s/Mpc for P18+BAO and $H_0=66.85^{+0.93}_{-0.84}$ km/s/Mpc for ACT+BAO.

When we analyze SN and BAO separately, the impact of different geometric measurements on our primary parameters of interest becomes apparent. Specifically, the BAO sample plays a crucial role in breaking the statistical degeneracy within our extensive parameter space. In conclusion, including BAO data, no room is left to solve the Hubble tension and the coupling parameter is very well limited.

\subsubsection{Joint Analyses}

We conclude by considering CMB, BAO, and (uncalibrated) SN data altogether in the analysis. In this case, which represents the most constraining dataset analyzed in the work, we show the joint constraints on $w_0$, $\xi$ and $H_0$ in \autoref{fig:1}. Taking the numerical results at face value, they read $w_0<-0.915$ and $\xi>-0.123$ for P18+SN+BAO. Instead, for ACT+SN+BAO, we get $w_0<-0.898$, $\xi>-0.110$, in very good agreement with the former. As largely expected from previous discussions on BAO and SN data, once we consider these combinations together, the constraints on the expansion rate of the Universe are very tight: $H_0=67.13\pm0.57$ km/s/Mpc for P18+SN+BAO and $H_0=66.72\pm0.61$ km/s/Mpc for ACT+SN+BAO. These values are in line with those derived within a standard cosmological model and therefore in $\sim 5\sigma$ tension with SH0ES.

\subsection{Phantom EoS}
\label{sec:pIDE}

We now turn to studying the phantom regime. The results obtained imposing a phantom EoS are provided in \autoref{tab.results.nondyanmical.phantom.P18} for the combinations of data involving the P18 CMB measurements and in \autoref{tab.results.nondyanmical.phantom.ACT} for the ACT data. As for the quintessence case, we consider CMB observations on their own and in different combinations involving SN and BAO distance measurements. In what follows, we summarize the most interesting findings.

\begin{table*}[htpb!]
\begin{center}
\renewcommand{\arraystretch}{1.5}
\resizebox{0.9 \textwidth}{!}{
\begin{tabular}{l c c c c c c c c c c c c c c c }
\hline
\textbf{Parameter} & \textbf{ P18 } & \textbf{ P18+SN } & \textbf{ P18+SN+SH0ES } & \textbf{ P18+BAO } & \textbf{ P18+BAO+SN } \\ 
\hline\hline

$ \Omega_\mathrm{b} h^2  $ & $  0.02242\pm 0.00014 $ & $  0.02233\pm 0.00014 $ & $0.02249\pm 0.00014$ & $  0.02236\pm 0.00014 $ & $  0.02237\pm 0.00013 $ \\
$ \Omega_\mathrm{c} h^2  $ & $  0.134^{+0.011}_{-0.012} $ & $  0.147\pm 0.011 $ &  $0.141\pm 0.013$ & $  0.1257^{+0.0043}_{-0.0054} $ & $  0.1241^{+0.0030}_{-0.0037} $ \\

$ 100\theta_\mathrm{s}  $ & $  1.04190\pm 0.00030 $ & $  1.04180\pm 0.00030 $ & $1.04200\pm 0.00029$ & $  1.04186\pm 0.00029 $ & $  1.04188\pm 0.00028 $ \\ 

$ \tau_\mathrm{reio}  $ & $  0.0544\pm 0.0066 $ & $  0.0538\pm 0.0076 $ & $0.0574\pm 0.0077$ & $  0.0543\pm 0.0073 $ & $  0.0568\pm 0.0073 $ \\ 

$ n_\mathrm{s}  $ & $  0.9662\pm 0.0042 $ & $  0.9632\pm 0.0042 $ & $0.9673\pm 0.0040$ & $  0.9644\pm 0.0039 $ & $  0.9650\pm 0.0037 $ \\ 

$ \log(10^{10} A_\mathrm{s})  $ & $  3.042\pm 0.013 $ & $  3.044\pm 0.015 $ & $3.048\pm 0.015$ & $  3.044\pm 0.014 $ & $  3.049\pm 0.014 $ \\ 

$ \xi  $ & $ < 0.297 $ & $  0.29^{+0.14}_{-0.17}\, (< 0.515 ) $ & $< 0.475$ & $ < 0.135 $ & $ < 0.0990 $ \\ 

$ w_0  $ & $ > -2.40 $ & $ > -1.16 $ & $-1.132^{+0.063}_{-0.052}\, (-1.13^{+0.11}_{-0.10})$ & $ > -1.20 $ & $ > -1.07 $ \\

$ H_0  $ [km/s/Mpc] & $  94\pm 20 $ & $  66.69\pm 0.82 $ & $69.82\pm 0.70$ & $  70.1\pm 1.2 $ & $  68.02\pm 0.52 $ \\ 

$ \Omega_\mathrm{m}  $ & $  0.196^{+0.069}_{-0.087} $ & $  0.382\pm 0.030 $ & $0.336\pm 0.029$ & $  0.303\pm 0.011 $ & $  0.3180\pm 0.0091 $ \\

$ \sigma_8  $ & $  0.92\pm 0.12 $ & $  0.690^{+0.049}_{-0.056} $ & $0.722\pm 0.058$ & $  0.805^{+0.024}_{-0.021} $ & $  0.794^{+0.020}_{-0.018} $ \\ 

$ r_\mathrm{drag}  $ & $  147.26\pm 0.27 $ & $  147.08\pm 0.26 $ & $147.32\pm 0.26$ & $  147.12\pm 0.24 $ & $  147.17\pm 0.22 $ \\ 

\hline \hline
\end{tabular} }
\end{center}
\caption{Constraints at 68\% (95\%) CL and upper limits at 95\% CL on the parameters of the $w_0$IDE model with $w_0<-1$. The results are obtained by different combinations of P18, BAO and SN (with and without the SH0ES calibration) measurements.}
\label{tab.results.nondyanmical.phantom.P18}
\end{table*}

\begin{table*}[htpb!]
\begin{center}
\renewcommand{\arraystretch}{1.5}
\resizebox{0.9 \textwidth}{!}{
\begin{tabular}{l c c c c c c c c c c c c c c c }
\hline
\textbf{Parameter} & \textbf{ ACT } & \textbf{ ACT+SN } & \textbf{ ACT+SN+SH0ES } & \textbf{ ACT+BAO } & \textbf{ ACT+BAO+SN } \\ 
\hline\hline

$ \Omega_\mathrm{b} h^2  $ & $0.02157\pm 0.00030$ & $  0.02156\pm 0.00030 $ & $0.02188\pm 0.00030$ & $  0.02158\pm 0.00030 $ & $  0.02155\pm 0.00030 $ \\ 

$ \Omega_\mathrm{c} h^2  $ & $0.1335^{+0.0070}_{-0.0144}$ & $  0.146\pm 0.011 $ & $0.1360^{+0.0085}_{-0.0167}$ & $  0.1262^{+0.0047}_{-0.0058} $ & $  0.1251^{+0.0035}_{-0.0041} $ \\ 

$ 100\theta_\mathrm{s}  $ & $1.04327\pm 0.00068$ & $  1.04314\pm 0.00066 $ & $1.04370\pm 0.00066$ & $  1.04313\pm 0.00064 $ & $  1.04324\pm 0.00065 $ \\ 

$ \tau_\mathrm{reio}  $ & $0.065\pm 0.014$ & $  0.066\pm 0.014 $ & $0.082\pm 0.014$ & $  0.066\pm 0.012 $ & $  0.071\pm 0.011 $ \\ 

$ n_\mathrm{s}  $ & $1.001\pm 0.012$ & $  0.996\pm 0.012 $ & $0.994\pm 0.017$ & $  0.996\pm 0.012 $ & $  0.998\pm 0.012 $ \\ 

$ \log(10^{10} A_\mathrm{s})  $ & $3.057^{+0.026}_{-0.025}$ & $  3.064\pm 0.025 $ & $3.094\pm 0.024$ & $  3.063\pm 0.022 $ & $  3.071\pm 0.021 $ \\ 

$ \xi  $ & $<0.150$ & $  0.28^{+0.13}_{-0.17}\, ( 0.28^{+0.25}_{-0.28} ) $ & $<0.438$ & $ < 0.149 $ & $ < 0.120 $ \\ 

$ w_0  $ & $-1.71^{+0.61}_{-0.30}\, (>-2.38)$ & $ > -1.15 $ & $-1.097^{+0.062}_{-0.036}\, (>-1.184)$ & $ > -1.18 $ & $ > -1.07 $ \\

$ H_0  $ [km/s/Mpc] & $91^{+11}_{-23}$ & $  66.39\pm 0.86 $ & $69.97\pm 0.69$ & $  69.2^{+1.1}_{-1.3} $ & $  67.64\pm 0.58 $ \\ 

$ \Omega_\mathrm{m}  $ & $0.208^{+0.042}_{-0.106}$ & $  0.382\pm 0.029 $ & $0.324^{+0.020}_{-0.034}$ & $  0.310\pm 0.012 $ & $  0.322\pm 0.011 $ \\ 

$ \sigma_8  $ & $0.93\pm 0.14$ & $  0.709\pm 0.048 $ & $0.746^{+0.071}_{-0.050}$ & $  0.817^{+0.026}_{-0.023} $ & $  0.809^{+0.023}_{-0.021} $ \\ 

$ r_\mathrm{drag}  $ & $148.39\pm 0.67$ & $  147.99\pm 0.62 $ & $148.61\pm 0.60$ & $  148.03\pm 0.50 $ & $  148.14\pm 0.48 $ \\ 

\hline \hline
\end{tabular} }
\end{center}
\caption{Constraints at 68\% (95\%) CL and upper limits at 95\% CL on the parameters of the $w_0$IDE model with $w_0<-1$. The results are obtained by different combinations of ACT, BAO and SN (with and without the SH0ES calibration) measurements.}
\label{tab.results.nondyanmical.phantom.ACT}
\end{table*}

\begin{figure*}[htpb!]
    \centering
    \includegraphics[width=0.7\linewidth]{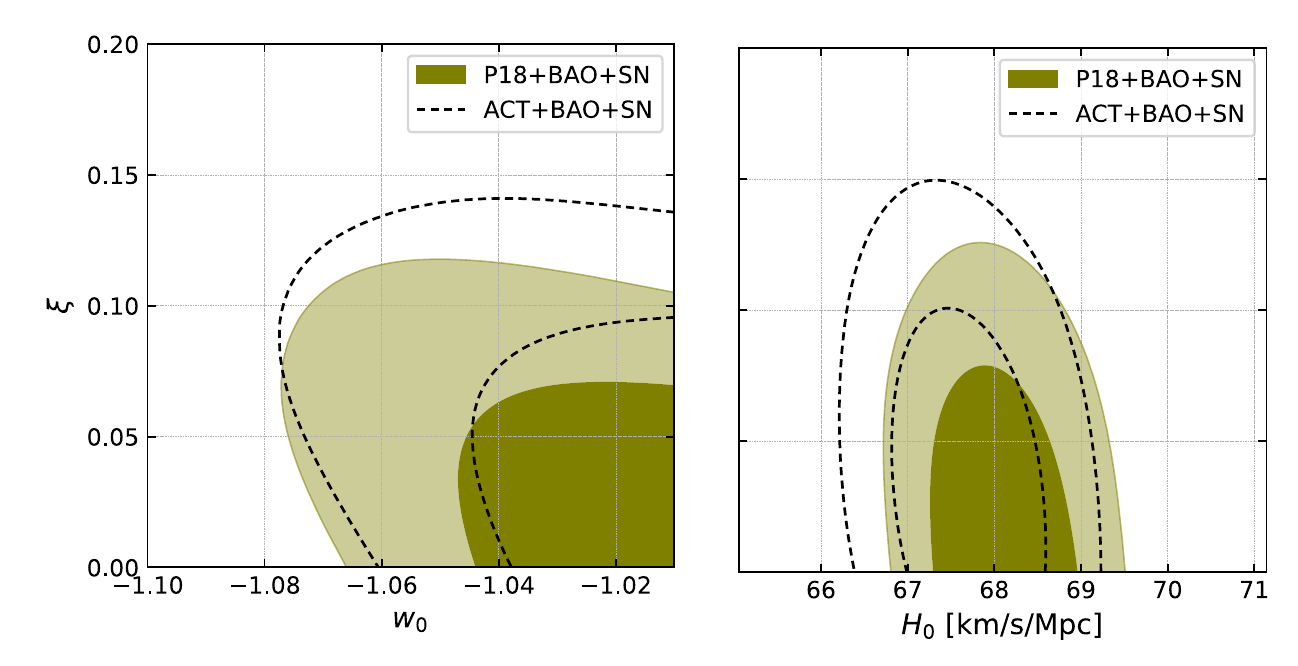}
    \caption{Joint marginalized contours at 68\% and 95\% CL illustrating the correlation between the coupling parameter $\xi$, the non-dynamical phantom EoS $w_0$, and the Hubble parameter $H_0$ for P18+BAO+SN and ACT+BAO+SN.}
    \label{fig:2}
\end{figure*}

\subsubsection{CMB-only}

As usual, geometrical degeneracy among different parameters strongly reduces the precision we can achieve from CMB data. In this case, from P18, we obtain $w_0 > -2.40$ and $\xi < 0.297$. Interestingly, for ACT, we find $w_0 > -2.38$ and $\xi < 0.150$. Therefore, we note that the well-known Planck preference for a phantom equation of state observed within the minimal extended $w_0$CDM model~\cite{Escamilla:2023oce} (i.e., with no energy-momentum transfer between dark matter and dark energy) here is reflected in the fact that P18 prefers a larger $\xi$ compared to ACT. That being said, we stress once more that both of these bounds are very large, confirming that focusing exclusively on CMB measurements is not ideal to constrain IDE when the DE EoS is allowed to vary in the model. This lack of constraining power mainly reflects on the results we can obtain for the present-day expansion rate, which is essentially unconstrained in both P18 ($H_0 = 94 \pm 20$ km/s/Mpc) and ACT ($H_0 = 91^{+11}_{-23}$ km/s/Mpc).

\subsubsection{CMB and SN}

As a next step, we introduce SN measurements from the \textit{Pantheon-Plus} catalogue. Following what has been done for quintessence models, we distinguish two different cases, presenting the results obtained with uncalibrated SN and SH0ES calibrated SN separately.

To begin with, we consider the uncalibrated dataset. In this case, the constraints on the DE EoS ($w_0 > -1.16$ for P18+SN and $w_0 > -1.15$ for ACT+SN) are significantly more constraining, ruling out a large portion of the parameter space allowed in the CMB-only case and narrowing down deviations from a cosmological constant to $\lesssim 15$\%. Interestingly, when we break the degeneracy between the different parameters, from P18+SN we get an indication at 68\% CL for a non-vanishing interaction, $\xi = 0.29^{+0.14}_{-0.17}$, which is supported by ACT+SN $\xi = 0.28^{+0.13}_{-0.17}$. However, in both cases, this indication is essentially lost at 95\% CL. Concerning the Hubble parameter, just like in the quintessence case, including uncalibrated SN leads to values of $H_0$ in tension with SH0ES: from P18+SN we get $H_0=66.69\pm0.82$ km/s/Mpc while from ACT+SN we get $H_0=66.39\pm0.86$ km/s/Mpc. As a result, also in the phantom case considering uncalibrated SN measurements we are unable to alleviate the Hubble tension.

Considering SN calibrated with SH0ES, from P18+SN+SH0ES we constrain $w = -1.132^{+0.063}_{-0.052}$. This is in good agreement with the result we get for ACT+SN+SH0ES: $w = -1.097^{+0.062}_{-0.036}$, both given at 68\% CL. Additionally, including the SH0ES calibration allows for a larger amount of energy-momentum to be transferred from DE to DM. This is evident from the upper limits on the coupling parameter ($\xi < 0.475$ for P18+SN+SH0ES and $\xi < 0.438$ for ACT+SN+SH0ES). As a result, we are now able to increase the present-day expansion rate of the Universe to $H_0 = 69.82 \pm 0.70$ km/s/Mpc and $H_0 = 69.97 \pm 0.69$ km/s/Mpc for P18+SN+SH0ES and ACT+SN+SH0ES, respectively. Therefore, in this case, the Hubble tension would be reduced down to $\sim 2.5 - 2.7\sigma$, just like in the quintessence case. However, in this case, the tension is reduced because of the effects of phantom EoS rather than because of interactions.

In conclusion, taking SN data at face value and focusing on phantom models, we reach the very same conclusion already pointed out for the quintessence regime. Regardless of whether the energy-momentum transfer flows from DM to DE or from DE to DM, if we consider uncalibrated supernovae data, IDE cannot represent a solution to the Hubble tension. However, we can mitigate the problem by calibrating this dataset with SH0ES.

\subsubsection{CMB and BAO}

We shall now consider CMB data in combination with BAO. In this case, we find a somewhat surprising outcome. First and foremost, we note that from P18+BAO, we get $w_0 > -1.20$, in good agreement with ACT+BAO, which gives $w_0 > -1.18$. As usual, BAO data strongly limit the total amount of energy transferred from DE to DM, resulting in very tight 95\% upper limits on the coupling parameter: $\xi < 0.135$ for P18+BAO and $\xi < 0.149$ for ACT+BAO. Notice also that the two CMB experiments agree quite well. Nevertheless, the real element of surprise is that in this case, we can fit CMB and BAO data while obtaining a value of $H_0$ in agreement with local distance ladder measurements. Indeed, from Planck+BAO, we get $H_0 = 70 \pm 1.2$ km/s/Mpc, and similarly from ACT+BAO, we have $H_0 = 69.2^{+1.1}_{-1.3}$ km/s/Mpc. This is the opposite behavior we observed in the quintessence case. It is also very different from the results we obtained analyzing uncalibrated SN. In this regard, the difference with respect to SN measurements is that BAO seems to prefer a smaller matter density parameter $\Omega_m$ compared to SN, resulting in a smaller amount of energy converted from DE to DM (i.e., into more stringent constraints on the coupling $\xi$) and  increasing $H_0$. 
	
In conclusion, based on CMB+BAO data, a minimal phantom $w_0$IDE cosmology could possibly represent a possible solution for the Hubble constant tension. 

\subsubsection{Joint Analyses}

As usual, we conclude by performing a joint analysis of CMB, BAO, and SN data, namely our most constraining dataset. The correlation among $\xi$, $w_0$ and $H_0$ are shown in \autoref{fig:2} for both P18+BAO+SN and ACT+BAO+SN. When we combine all these data together, we become very restrictive on the DE EoS. Essentially, both P18+BAO+SN and ACT+BAO+SN analyses yield $w_0 > -1.07$ at a 95\% CL. This limit reduces our freedom to consider deviations away from a value resembling a cosmological constant to less than 7\%. Similarly, we become very constrained on the coupling between DM and DE, limiting $\xi < 0.0990$ for P18+BAO+SN and $\xi < 0.120$ for ACT+BAO+SN. Concerning the value of the present-day expansion rate, from P18+BAO+SN, we have $H_0 = 68.02 \pm 0.52$ km/s/Mpc, while from ACT+BAO+SN, we get $H_0 = 67.64 \pm 0.58$ km/s/Mpc. 
Therefore, combining BAO and SN data together, we lose the ability to increase the expansion rate of the Universe observed in the CMB(+BAO/SN+SH0ES) analyses, see also \autoref{fig:2}. Essentially, we obtain values of $H_0$ that, while larger than what is obtained within a minimal $\Lambda$CDM model of cosmology, remain in strong tension with local measurements from the SH0ES collaboration at $\sim 4\sigma$.

\bigskip

We conclude this section with an important final remark: the cases $w_0 > -1$ and $w_0 < -1$ do not necessarily produce the same magnitudes of the coupling parameter $\xi$, nor the same value of $H_0$ in the limit $\xi \to 0$. These discrepancies arise because the sign of $\xi$ induces different corrections in $H_0$ and $\Omega_m$, which combine with the well-known correlation between $w_0$ and $H_0$, differing in the quintessence and phantom regimes. As a result, due to the varying correlations introduced by $\xi$, the final outcomes of analyses with $w_0 > -1$ and $w_0 < -1$ should not be expected to match. However, from a physical point of view, this is not a cause for concern because the two regimes differ significantly in their physical nature and potential microphysical realizations. The sign of $\xi$ predicts distinct cosmological behaviors, featuring an energy-momentum flow in opposite directions, and this could fundamentally alter the model's physical and theoretical characteristics.
\clearpage
\section{Results for Dynamical EoS}
\label{sec:Results_DDE}

In this section, we discuss the results for a Dynamical DE EoS given by the CPL parameterization in Eq.~\eqref{eq:CPL}. We divide the section into two different subsections. In \autoref{sec:qDIDE}, we focus on quintessence models characterized by an EoS $w(z) > -1$ at any $z$ and a DM-to-DE energy-momentum flow (i.e., $\xi < 0$). Instead, in \autoref{sec:pDIDE}, we study phantom models with $w(z) < -1$ at any $z$ and a DE-to-DM energy-momentum transfer ($\xi > 0$).

\subsection{Quintessence EoS}
\label{sec:qDIDE}

The results obtained imposing a quintessence dynamical DE EoS are provided in \autoref{tab.results.dyanmical.quintessence.P18} for the combinations of data involving P18 and in \autoref{tab.results.dyanmical.quintessence.ACT} for those involving ACT. 

\begin{table*}[htbp!]
\begin{center}
\renewcommand{\arraystretch}{1.5}
\resizebox{0.9 \textwidth}{!}{
\begin{tabular}{l c c c c c c c c c c c c c c c }
\hline
\textbf{Parameter} & \textbf{ P18 } & \textbf{ P18+SN } & \textbf{ P18+SN+SH0ES } & \textbf{ P18+BAO } & \textbf{ P18+BAO+SN } \\ 
\hline\hline

$ \Omega_\mathrm{b} h^2  $ & $  0.02239\pm 0.00015 $ & $  0.02230\pm 0.00014 $ & $  0.02251\pm 0.00014 $ & $  0.02242\pm 0.00014 $ & $  0.02238\pm 0.00014 $ \\ 
$ \Omega_\mathrm{c} h^2  $ & $  0.068\pm 0.033 $ & $  0.1157^{+0.0052}_{-0.0039} $ & $  0.0987\pm 0.0079 $ & $  0.1142^{+0.0036}_{-0.0031} $ & $  0.1160^{+0.0029}_{-0.0025} $ \\ 
$ 100\theta_\mathrm{s}  $ & $  1.04192\pm 0.00030 $ & $  1.04180\pm 0.00030 $ & $  1.04205\pm 0.00029 $ & $  1.04192\pm 0.00028 $ & $  1.04190\pm 0.00028 $ \\ 
$ \tau_\mathrm{reio}  $ & $  0.0531\pm 0.0072 $ & $  0.0522\pm 0.0073 $ & $  0.0577\pm 0.0077 $ & $  0.0564\pm 0.0073 $ & $  0.0558\pm 0.0072 $ \\ 
$ n_\mathrm{s}  $ & $  0.9661\pm 0.0043 $ & $  0.9627\pm 0.0040 $ & $  0.9684\pm 0.0040 $ & $  0.9663\pm 0.0038 $ & $  0.9653\pm 0.0037 $ \\ 
$ \log(10^{10} A_\mathrm{s})  $ & $  3.041\pm 0.014 $ & $  3.041\pm 0.014 $ & $  3.048\pm 0.015 $ & $  3.047\pm 0.014 $ & $  3.047\pm 0.014 $ \\ 

$ \xi  $ & $  -0.40^{+0.36}_{-0.15}\, (> -0.766 ) $ & $ > -0.126  $ & $  -0.171^{+0.082}_{-0.070}\, ( -0.17^{+0.15}_{-0.14} ) $ & $  -0.046^{+0.044}_{-0.013}\, (> -0.0970 ) $ & $ > -0.0781  $ \\ 

$ w_{a}  $ & $ > -1.22 $ & $ > -1.21 $ & $ > -1.25 $ & $ > -1.24 $ & $ > -1.19 $ \\ 
$ w_0  $ & $  --- $ & $  --- $ & $  --- $ & $  --- $ & $  --- $ \\ 
$ H_0  $ [km/s/Mpc] & $  71.6\pm 2.5 $ & $  67.50\pm 0.60 $ & $  69.81\pm 0.65 $ & $  68.16\pm 0.48 $ & $  67.86\pm 0.43 $ \\ 
$ \Omega_\mathrm{m}  $ & $  0.182\pm 0.077 $ & $  0.304^{+0.016}_{-0.013} $ & $  0.250\pm 0.020 $ & $  0.2956\pm 0.0099 $ & $  0.3019\pm 0.0084 $ \\ 
$ \sigma_8  $ & $  1.63^{+0.55}_{-1.3} $ & $  0.845^{+0.027}_{-0.038} $ & $  0.959^{+0.074}_{-0.085} $ & $  0.843^{+0.021}_{-0.024} $ & $  0.835^{+0.018}_{-0.021} $ \\ 
$ r_\mathrm{drag}  $ & $  147.18\pm 0.27 $ & $  146.99\pm 0.26 $ & $  147.33\pm 0.26 $ & $  147.25\pm 0.23 $ & $  147.18\pm 0.22 $ \\ 

\hline \hline
\end{tabular} }
\end{center}
\caption{Constraints at 68\% (95\%) CL and upper limits at 95\% CL on the parameters of the dynamical $w(z)$IDE model with $w(z)>-1$ at any $z$. The results are obtained by different combinations of P18, BAO and SN (with and without the SH0ES calibration) measurements.}
\label{tab.results.dyanmical.quintessence.P18}
\end{table*}

\begin{table*}[htbp!]
\begin{center}
\renewcommand{\arraystretch}{1.5}
\resizebox{0.9 \textwidth}{!}{
\begin{tabular}{l c c c c c c c c c c c c c c c }
\hline
\textbf{Parameter} & \textbf{ ACT } & \textbf{ ACT+SN } & \textbf{ ACT+SN+SH0ES } & \textbf{ ACT+BAO } & \textbf{ ACT+BAO+SN } \\ 
\hline\hline

$ \Omega_\mathrm{b} h^2  $ & $  0.02161\pm 0.00030 $ & $  0.02153\pm 0.00029 $ & $  0.02188\pm 0.00030 $ & $  0.02163\pm 0.00030 $ & $  0.02159\pm 0.00029 $ \\ 
$ \Omega_\mathrm{c} h^2  $ & $ < 0.0713 $ & $  0.1158^{+0.0056}_{-0.0043} $ & $  0.1017^{+0.0088}_{-0.0079} $ & $  0.1154^{+0.0031}_{-0.0026} $ & $  0.1167^{+0.0025}_{-0.0022} $ \\ 
$ 100\theta_\mathrm{s}  $ & $  1.04329\pm 0.00066 $ & $  1.04303\pm 0.00065 $ & $  1.04370\pm 0.00063 $ & $  1.04320\pm 0.00063 $ & $  1.04317\pm 0.00062 $ \\ 
$ \tau_\mathrm{reio}  $ & $  0.067\pm 0.014 $ & $  0.059\pm 0.013 $ & $  0.080\pm 0.013 $ & $  0.070\pm 0.011 $ & $  0.069\pm 0.011 $ \\ 
$ n_\mathrm{s}  $ & $  0.999\pm 0.012 $ & $  0.995\pm 0.012 $ & $  0.995\pm 0.012 $ & $  0.996\pm 0.012 $ & $  0.996\pm 0.012 $ \\ 
$ \log(10^{10} A_\mathrm{s})  $ & $  3.063\pm 0.025 $ & $  3.050\pm 0.024 $ & $  3.091\pm 0.023 $ & $  3.071\pm 0.021 $ & $  3.069\pm 0.020 $ \\ 

$ \xi  $ & $  -0.49^{+0.23}_{-0.28}\, ( -0.49^{+0.49}_{-0.36} ) $ & $ > -0.146 $ & $  -0.127^{+0.10}_{-0.049}\, (> -0.252 ) $ & $ > -0.0779 $ & $ > -0.0642 $ \\ 

$ w_{a}  $ & $ > -1.21 $ & $ > -1.22 $ & $ > -1.22 $ & $ > -1.29 $ & $ > -1.16 $ \\ 
$ w_0  $ & $  --- $ & $  --- $ & $  --- $ & $  --- $ & $  --- $ \\ 
$ H_0  $ [km/s/Mpc] & $  72.6^{+3.1}_{-2.7} $ & $  67.07\pm 0.78 $ & $  70.08\pm 0.68 $ & $  67.87\pm 0.54 $ & $  67.57\pm 0.48 $ \\ 
$ \Omega_\mathrm{m}  $ & $  0.148^{+0.078}_{-0.095} $ & $  0.307^{+0.018}_{-0.016} $ & $  0.253\pm 0.019 $ & $  0.2990\pm 0.0094 $ & $  0.3045\pm 0.0079 $ \\ 
$ \sigma_8  $ & $  2.10^{+0.91}_{-1.6} $ & $  0.869^{+0.031}_{-0.043} $ & $  0.947^{+0.072}_{-0.086} $ & $  0.856^{+0.018}_{-0.022} $ & $  0.851^{+0.015}_{-0.018} $ \\ 
$ r_\mathrm{drag}  $ & $  148.30\pm 0.64 $ & $  147.71\pm 0.59 $ & $  148.55\pm 0.58 $ & $  148.22\pm 0.48 $ & $  148.12\pm 0.47 $ \\ 

\hline \hline
\end{tabular} }
\end{center}
\caption{Constraints at 68\% (95\%) CL and upper limits at 95\% CL on the parameters of the dynamical $w(z)$IDE model with $w(z)>-1$ at any $z$. The results are obtained by different combinations of ACT, BAO and SN (with and without the SH0ES calibration) measurements.}
\label{tab.results.dyanmical.quintessence.ACT}
\end{table*}

\begin{figure*}[ht!]
    \centering
    \includegraphics[width=0.9\linewidth]{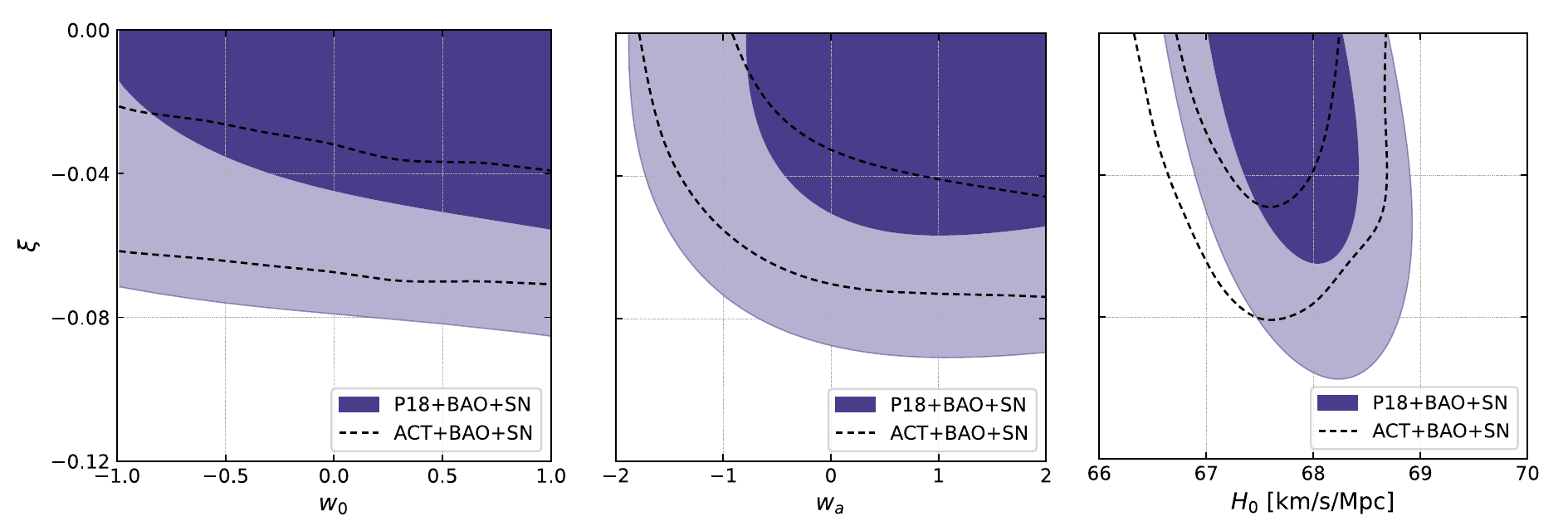}
    \caption{Joint marginalized contours at 68\% and 95\% CL illustrating the correlation between the coupling parameter $\xi$, the dynamical EoS parameters $w_0$ and $w_a$ of the CPL parameterization (obtained by imposing a quintessence EoS $w(z)>-1$ for any $z$), and the Hubble parameter $H_0$ for P18+BAO+SN and ACT+BAO+SN.}
    \label{fig:3}
\end{figure*}

\subsubsection{CMB-only}

First and foremost, we note that all the concerns we pointed out in the non-dynamical case about the geometrical degeneracy among cosmological parameters observed for the CMB-only case apply also in the dynamical case. This problem can become even more relevant since we now have one additional parameter, $w_a$, describing the redshift evolution of $w(z)$. With this premise, it is not surprising that for P18 we are not able to constrain $w_0$.\footnote{Given that the constraints on the DE EoS often reach the prior bounds, a few important remarks on the prior range adopted for $w_0$ and $w_a$ and their possible implications for constraints on relevant parameters such as $H_0$ and $\xi$ are in order. Firstly, we note that no significant correlation is found between the parameters describing the EoS, $H_0$, and $\xi$. As seen from \autoref{fig:3} and ~\autoref{fig:4}, the probability contours at 68\% and 95\% are essentially flat and do not show significant shifts in $\xi$ and $H_0$ when $w_0$ and $w_a$ are varied. Therefore, the choice of prior on the DE EoS is not crucial for the constraints on key parameters related to cosmological tensions. We explicitly tested this aspect by adopting different case study priors on $w_0$ and $w_a$ without observing any relevant changes in the results for $H_0$ and $\xi$ (in both quintessence and phantom cases). Notice also that, for direct comparison of the results, we are using the standard priors on all parameters that are widely adopted in the literature.} However, we can get a 95\% lower limit on $w_a > -1.22$. Similar results are obtained for ACT: $w_0$ is unconstrained while $w_a$ turns out to be $w_a > -1.21$ at 95\% CL. Concerning the coupling parameter, we note that in the dynamical case, we get a preference for a non-vanishing interaction $\xi = -0.40^{+0.36}_{-0.15}$ for P18 and $\xi = -0.49^{+0.23}_{-0.28}$ for ACT. Also in the dynamical case, the two experiments agree about the possible amount of energy-momentum to be transferred from DM to DE. This preference towards an interacting dark sector produces higher values of the present-day expansion rate which reads $H_0 = 71.6 \pm 2.5$ km/s/Mpc for P18 and $H_0 = 72.6^{+3.1}_{-2.7}$ km/s/Mpc for ACT. Therefore, despite having large uncertainties, the CMB-only case suggests that considering dynamical quintessence models can facilitate solving the Hubble tension compared to the respective non-dynamical case. However, to confirm this preference, it is mandatory to test the model against low-redshift data.

\subsubsection{CMB and SN}

As usual, we consider both uncalibrated and SH0ES calibrated SN data. The first thing we stress is that including low-redshfit observations does not significantly improve the constraints on the EoS parameters, $w_0$ and $w_a$. In this extended model the EoS can change over time while allowing energy-momentum exchange. Due to the large number of free degrees of freedom, data do not have enough power to constrain all the parameters simultaneously.

Despite not being able to say much about the DE EoS, considering uncalibrated SNe significantly increases the constraining power on $\xi$. Specifically, we lose the preference for interactions, obtaining $\xi > -0.126$ from P18+SN and $\xi > -0.146$ from ACT+SN, both at 95\% CL. As a result, we recover values of the Hubble parameter in line with a baseline $\Lambda$CDM cosmology, and in tension with SH0ES ($H_0 = 67.50 \pm 0.60$ km/s/Mpc for P18+SN and $H_0 = 67.07 \pm 0.78$ km/s/Mpc for ACT+SN).

Following the other option on the table, we consider the \textit{Pantheon-Plus} catalogue calibrated with SH0ES. We stress again that this possibility is well motivated as the CMB-only analysis, despite large uncertainties, suggests a significant shift towards higher fitting values of $H_0$. From P18+SN+SH0ES ($\xi = -0.171^{+0.082}_{-0.070}$) and ACT+SN+SH0ES ($\xi = -0.127^{+0.10}_{-0.049}$), we find a mild preference for a non-vanishing interaction. However, these results are much more constraining about the total amount of energy-momentum transfer allowed in the model, if compared with the CMB-only case. This somehow reduces the model's ability to predict higher values of the present-day expansion rate ($H_0 = 69.81 \pm 0.65$ km/s/Mpc for P18+SN+SH0ES and $H_0 = 70.08 \pm 0.68$ km/s/Mpc for ACT+SN+SH0ES). Despite this limitation, quintessence dynamical models are still able to reduce the $H_0$-tension to $2.4 - 2.7\sigma$.

As a result, for the dynamical quintessence case, we can derive the same conclusions obtained for the non-dynamical quintessence model: combining CMB with uncalibrated SN the model can hardly be considered a possible solution to the Hubble tension. However, using the SH0ES calibration leaves enough room to mitigate (but not completely solve) the problem.

\subsubsection{CMB and BAO}

As a next step, we consider CMB+BAO. In this case, we become very restrictive on the coupling $\xi$. From P18+BAO, we still get a very tiny preference for $\xi=-0.046^{+0.044}_{-0.013}$. However, this preference is lost at 95\% CL and is not confirmed by ACT ($\xi>-0.0779$). Concerning the Hubble rate, from P18+BAO, we obtain $H_0=68.16\pm0.48$ km/s/Mpc. Similarly, for ACT, we get $H_0=67.87\pm0.54$ km/s/Mpc. Both these values are larger than the respective results in $\Lambda$CDM cosmology, yet in tension with SH0ES at more than $4\sigma$. 

As a result, one more time BAO data do not leave room to solve the Hubble tension in the contest of IDE, not  even allowing for a dynamical quintessence EoS.

\subsubsection{Joint Analyses}

We conclude the study of the dynamical quintessence model by analyzing CMB, SN, and BAO data together. In this case, the 2D correlations between the coupling $\xi$, the EoS parameters $w_0$ and $w_a$, and the Hubble rate $H_0$ are shown in \autoref{fig:3}. From the figure, it is evident that even for our most constraining dataset, we do not have enough power to narrow down the parameter space allowed for $w_0$ and $w_a$ that shape the redshift behavior of $w(z)$. In contrast, we are very restrictive on both $\xi$ and $H_0$. The numerical results read $\xi > -0.0781$ and $H_0 = 67.86 \pm 0.43$ km/s/Mpc for P18+BAO+SN, while for ACT+BAO+SN we have $\xi > -0.0642$ and $H_0 = 67.57 \pm 0.48$ km/s/Mpc. Again, everything is in agreement with a baseline $\Lambda$CDM cosmology.

\subsection{Phantom EoS}
\label{sec:pDIDE}

We conclude our explorative study of IDE cosmology by considering models featuring a phantom dynamical EoS $w(z) < -1$. The results involving P18 are given in \autoref{tab.results.dynamical.phantom.P18}, those involving ACT are given in \autoref{tab.results.dynamical.phantom.ACT}.

\begin{table*}[htbp!]
\begin{center}
\renewcommand{\arraystretch}{1.5}
\resizebox{0.9 \textwidth}{!}{
\begin{tabular}{l c c c c c c c c c c c c c c c }
\hline
\textbf{Parameter} & \textbf{ P18 } & \textbf{ P18+SN } & \textbf{ P18+SN+SH0ES } & \textbf{ P18+BAO } & \textbf{ P18+BAO+SN } \\ 
\hline\hline

$ \Omega_\mathrm{b} h^2  $ & $  0.02232\pm 0.00015 $ & $  0.02235\pm 0.00014 $ & $  0.02263\pm 0.00014 $ & $  0.02243\pm 0.00013 $ & $  0.02241\pm 0.00013 $ \\ 
$ \Omega_\mathrm{c} h^2  $ & $  0.140^{+0.014}_{-0.015} $ & $  0.1302^{+0.0061}_{-0.0069} $ & $  0.1197^{+0.0022}_{-0.0027} $ & $  0.1212^{+0.0020}_{-0.0024} $ & $  0.1218^{+0.0021}_{-0.0025} $ \\ 
$ 100\theta_\mathrm{s}  $ & $  1.04181\pm 0.00029 $ & $  1.04184\pm 0.00030 $ & $  1.04217\pm 0.00028 $ & $  1.04196\pm 0.00029 $ & $  1.04193\pm 0.00028 $ \\ 
$ \tau_\mathrm{reio}  $ & $  0.0543\pm 0.0076 $ & $  0.0548\pm 0.0076 $ & $  0.0633\pm 0.0082 $ & $  0.0594\pm 0.0074 $ & $  0.0587\pm 0.0074 $ \\ 
$ n_\mathrm{s}  $ & $  0.9632\pm 0.0042 $ & $  0.9639\pm 0.0040 $ & $  0.9716\pm 0.0039 $ & $  0.9671\pm 0.0037 $ & $  0.9664\pm 0.0037 $ \\ 
$ \log(10^{10} A_\mathrm{s})  $ & $  3.045\pm 0.015 $ & $  3.045\pm 0.015 $ & $  3.057\pm 0.016 $ & $  3.053\pm 0.014 $ & $  3.052\pm 0.015 $ \\ 

$ \xi  $ & $ < 0.522 $ & $ < 0.224 $ & $ < 0.0647 $ & $ < 0.0583 $ & $ < 0.0642 $ \\

$ w_{a}  $ & $  --- $ & $ < 1.11 $ & $ < 1.12 $ & $ < 1.05 $ & $ < 1.10 $ \\ 
$ w_0  $ & $  --- $ & $  --- $ & $  --- $ & $  --- $ & $  --- $ \\ 
$ H_0  $ [km/s/Mpc] & $  65.1^{+1.9}_{-1.7} $ & $  66.31\pm 0.72 $ & $  68.48\pm 0.48 $ & $  67.62\pm 0.44 $ & $  67.47\pm 0.42 $ \\ 
$ \Omega_\mathrm{m}  $ & $  0.387^{+0.051}_{-0.061} $ & $  0.349^{+0.021}_{-0.024} $ & $  0.3049^{+0.0084}_{-0.0094} $ & $  0.3156^{+0.0079}_{-0.0089} $ & $  0.3182^{+0.0079}_{-0.0090} $ \\ 
$ \sigma_8  $ & $  0.707\pm 0.061 $ & $  0.753\pm 0.034 $ & $  0.792^{+0.018}_{-0.014} $ & $  0.797^{+0.015}_{-0.012} $ & $  0.796^{+0.015}_{-0.013} $ \\ 
$ r_\mathrm{drag}  $ & $  147.07\pm 0.27 $ & $  147.11\pm 0.26 $ & $  147.57\pm 0.25 $ & $  147.31\pm 0.23 $ & $  147.26\pm 0.22 $ \\ 

\hline \hline
\end{tabular} }
\end{center}
\caption{Constraints at 68\% (95\%) CL and upper limits at 95\% CL on the parameters of the dynamical $w(z)$IDE model with $w(z)<-1$ at any $z$. The results are obtained by different combinations of P18, BAO and SN (with and without the SH0ES calibration) measurements.}
\label{tab.results.dynamical.phantom.P18}
\end{table*}

\begin{table*}[htbp!]
\begin{center}
\renewcommand{\arraystretch}{1.5}
\resizebox{0.9 \textwidth}{!}{
\begin{tabular}{l c c c c c c c c c c c c c c c }
\hline
\textbf{Parameter} & \textbf{ ACT } & \textbf{ ACT+SN } & \textbf{ ACT+SN+SH0ES } & \textbf{ ACT+BAO } & \textbf{ ACT+BAO+SN } \\ 
\hline\hline

$ \Omega_\mathrm{b} h^2  $ & $  0.02156\pm 0.00030 $ & $  0.02160\pm 0.00030 $ & $  0.02197\pm 0.00029 $ & $  0.02162\pm 0.00029 $ & $  0.02158\pm 0.00029 $ \\ 
$ \Omega_\mathrm{c} h^2  $ & $  0.139\pm 0.013 $ & $  0.1308^{+0.0069}_{-0.0082} $ & $  0.1185^{+0.0033}_{-0.0041} $ & $  0.1215^{+0.0025}_{-0.0029} $ & $  0.1225^{+0.0026}_{-0.0031} $ \\ 
$ 100\theta_\mathrm{s}  $ & $  1.04317\pm 0.00066 $ & $  1.04315\pm 0.00066 $ & $  1.04394\pm 0.00063 $ & $  1.04328\pm 0.00063 $ & $  1.04330\pm 0.00063 $ \\ 
$ \tau_\mathrm{reio}  $ & $  0.069\pm 0.014 $ & $  0.069\pm 0.014 $ & $  0.093\pm 0.013 $ & $  0.076\pm 0.011 $ & $  0.075\pm 0.011 $ \\ 
$ n_\mathrm{s}  $ & $  0.995\pm 0.012 $ & $  0.994\pm 0.012 $ & $  0.995\pm 0.012 $ & $  0.998\pm 0.012 $ & $  0.998\pm 0.012 $ \\ 
$ \log(10^{10} A_\mathrm{s})  $ & $  3.069\pm 0.025 $ & $  3.071\pm 0.024 $ & $  3.116\pm 0.023 $ & $  3.081\pm 0.019 $ & $  3.079\pm 0.020 $ \\ 

$ \xi  $ & $ < 0.499 $ & $ < 0.269 $ & $ < 0.103 $ & $ < 0.0748 $ & $ < 0.0864 $ \\

$ w_{a}  $ & $ < 1.11 $ & $  --- $ & $ < 1.12 $ & $  --- $ & $  --- $ \\ 
$ w_0  $ & $  --- $ & $  --- $ & $  --- $ & $  --- $ & $  --- $ \\ 
$ H_0  $ [km/s/Mpc] & $  65.1\pm 1.7 $ & $  66.14\pm 0.89 $ & $  69.39\pm 0.63 $ & $  67.44\pm 0.53 $ & $  67.21\pm 0.50 $ \\ 
$ \Omega_\mathrm{m}  $ & $  0.384^{+0.052}_{-0.060} $ & $  0.350^{+0.024}_{-0.027} $ & $  0.293^{+0.011}_{-0.013} $ & $  0.3161\pm 0.0094 $ & $  0.3205\pm 0.0093 $ \\ 
$ \sigma_8  $ & $  0.728\pm 0.062 $ & $  0.767^{+0.049}_{-0.042} $ & $  0.806^{+0.028}_{-0.022} $ & $  0.815^{+0.019}_{-0.016} $ & $  0.812^{+0.020}_{-0.018} $ \\ 
$ r_\mathrm{drag}  $ & $  148.05\pm 0.63 $ & $  148.09\pm 0.59 $ & $  149.09\pm 0.56 $ & $  148.37\pm 0.46 $ & $  148.31\pm 0.47 $ \\ 

\hline \hline
\end{tabular} }
\end{center}
\caption{Constraints at 68\% (95\%) CL and upper limits at 95\% CL on the parameters of the dynamical $w(z)$IDE model with $w(z)<-1$ at any $z$. The results are obtained by different combinations of ACT, BAO and SN (with and without the SH0ES calibration) measurements.}
\label{tab.results.dynamical.phantom.ACT}
\end{table*}

\begin{figure*}[ht!]
    \centering
    \includegraphics[width=0.9\linewidth]{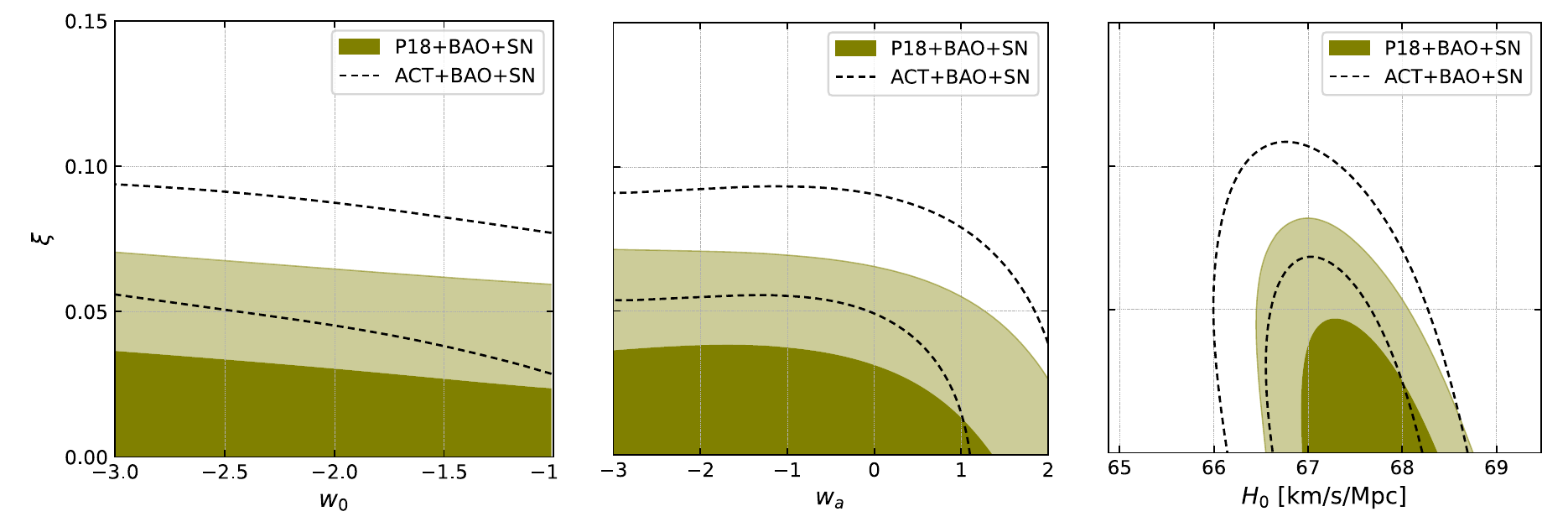}
    \caption{Joint marginalized contours at 68\% and 95\% CL illustrating the correlation between the coupling parameter $\xi$, the dynamical EoS parameters $w_0$ and $w_a$ of the CPL parameterization (obtained by imposing a phantom EoS $w(z)<-1$ for any $z$), and the Hubble parameter $H_0$ for P18+BAO+SN and ACT+BAO+SN.}
    \label{fig:4}
\end{figure*}

\subsubsection{CMB-only}

As usual, we start by considering only CMB temperature, polarization and lensing observations. Comparing the results with those derived for the non-dynamical case, we note that considering a dynamical phantom EoS allows a much larger fraction of energy-momentum flow from DE to DM. This is because within a CPL parametrization, we have more parameters to constrain. Taking the results at face value, from P18 we obtain $\xi < 0.522$, while for ACT we obtain $\xi < 0.499$. These large values of $\xi$ result in a preference for a larger matter fraction (fed by DE) and consequently into a shift in the Hubble parameter towards values smaller than the $\Lambda$CDM one ($H_0 = 65.1^{+1.9}_{-1.7}$ km/s/Mpc for P18 and $H_0 = 65.1 \pm 1.7$ km/s/Mpc for ACT). Therefore, dynamical phantom models may actually prove to be less effective in potentially addressing the Hubble constant tension compared to their non-dynamical counterpart. Having said that, as usual, CMB-only data leaves us with very large error bars, making the inclusion of additional local universe probes a necessary step to take before deriving any definitive conclusions.

\subsubsection{CMB and SN}

Combining CMB and SN together, we enhance the constraining power, narrowing down the upper limit on the coupling parameter $\xi$. Specifically, using uncalibrated SN, for P18+SN we obtain $\xi<0.224$, while for ACT+SN, we get $\xi<0.269$. Further tightening occurs when considering the SH0ES calibration. For P18+SN+SH0ES and ACT+SN+SH0ES, we find $\xi<0.0647$ and $\xi<0.103$, respectively. However, uncalibrated SN data yield lower values $H_0=66.31\pm0.72$ km/s/Mpc for P18+SN and $H_0=66.14 \pm 0.89$ km/s/Mpc for ACT+SN. These results are in clear tension with SH0ES. 

Overall, SN confirm that dynamical phantom models struggle to represent a possible solution to the $H_0$-tension, in line with what we anticipated from the CMB-only analysis.

\subsubsection{CMB and BAO}

When it comes to considering BAO data, we find that the coupling parameter is constrained to $\xi<0.0583$ for P18+BAO and $\xi<0.0748$ for ACT+BAO. Both datasets predict values of $H_0$ that are essentially the same obtained for $\Lambda$CDM.

\subsubsection{Joint Analyses}

Finally, we come to the most constraining CMB+SN+BAO dataset. The correlations among the most relevant parameters are shown in \autoref{fig:4}. As evident from the figure, the joint analysis confirms the overall trend observed consistently in the phantom dynamical case. The amount of energy and momentum that can be transferred from DE to DM is very constrained for both P18+BAO+SN ($\xi<0.0642$) and ACT+BAO+SN ($\xi<0.0864$). Everything is in line with a late-time $\Lambda$CDM cosmology, including the value inferred for the Hubble parameter.

\newpage
\section{Conclusions}
\label{sec:conclusions}

In this paper, we undertake a comprehensive reassessment of the constraints on IDE cosmology, namely cosmological models featuring energy-momentum flow between DM and DE.

Our model is detailed in \autoref{sec:theory}. On top of this model, we expand the dark sector physics, allowing for more freedom in the DE sector by not restricting the EoS to being that of a cosmological constant. We review, update and extend the state-of-the-art analyses performed in earlier similar studies by considering two distinct physical scenarios: IDE cosmology with a non-dynamical EoS $w_0\ne-1$, and IDE models with a dynamical EoS $w(z)$. For the latter, we adopt a simple CPL parameterization given by Eq.~\eqref{eq:CPL}.

Avoiding early-time superhorizon instabilities in the dynamics of cosmological perturbations imposes stability conditions on the DM-DE coupling $\xi$ and the DE EoS, forcing the fraction $\xi/(1+w)$ to be positive. Both in the dynamical and non-dynamical case, we have carefully taken into account stability conditions, studying separately two physical regimes represented by phantom and quintessence EoS. In the quintessence and phantom regimes, the energy-momentum transfer is forced to flow in different directions (from DM to DE and from DE to DM, respectively), producing a quite different phenomenology both in terms of perturbations and background dynamics.

Aimed to conclusively assess whether IDE models featuring dynamical and/or non-dynamical DE EoS can represent a possible solution to the Hubble constant tension, we systematically study all the possibilities deriving updated observational constraints from the latest cosmological and astrophysical observations. Specifically, we consider two different independent CMB experiments: the Planck-2018 temperature polarization and lensing data as well as small-scale Atacama Cosmology Telescope CMB measurements. CMB experiments are considered on their own as well as in different combinations involving low-redshift probes, such as Supernovae distance moduli measurements from the \textit{Pantheon-Plus} catalog and the most recent Baryon Acoustic Oscillations from the SDSS-IV eBOSS survey.

Our updated and extended analysis reveals significant differences compared to the state-of-the-art results, significantly restricting the parameter space allowed to IDE models with dynamical and non-dynamical EoS, as well as limiting their overall ability to reconcile cosmological tensions. Notably, all our most important findings are always independently corroborated by the two different CMB experiments (that share a consistency of view on IDE even allowing the dark sector physics), making the conclusions of our analysis robust. The most important takeaway results read as follows:

\begin{itemize}[leftmargin=*]

\item IDE models featuring a non-dynamical quintessence EoS ($w_0 > -1$) produce larger values of the present-day expansion rate $H_0$ when analyzed in terms of CMB data. This is due to the fact that the total amount of energy-momentum flow allowed from the DM to the DE is poorly constrained, leaving enough freedom to obtain large negative $\xi$ and higher $H_0$. However, including low-redshift probes, this preference is essentially lost. While using SN measurements calibrated with SH0ES, we can still obtain a value of $H_0$ large enough to reduce the Hubble tension down to $2.5 - 2.7 \sigma$. Considering the uncalibrated SN dataset or BAO distance measurements (both separately and in conjunction), we become very constrained on the coupling $\xi$, and no room is left to increase $H_0$ towards local distance ladder values anymore. The most constraining cosmological bounds on these scenarios are summarized in \autoref{fig:1}.

\item IDE models featuring a non-dynamical phantom EoS ($w_0 < -1$) predict an energy-momentum transfer from the DE to DM. Interestingly, when this model is analyzed with CMB and CMB+BAO, and CMB+SN+SH0ES data, we get larger values of the Hubble constant, primarily due to a phantom DE EoS. Therefore, this model can, in principle, help with the Hubble tension as well. However, considering uncalibrated SN (both in combination CMB+SN and CMB+BAO+SN), the preference for larger $H_0$ is strongly reduced. Also, in this case, the joint analysis of CMB, SN, and BAO data (whose results are shown in \autoref{fig:2}) strongly limits the ability of the model to represent a solution to the Hubble tension.

\item IDE models featuring a dynamical quintessence EoS ($w(z) > -1$ at any $z$) perform better in attempting to increase the value of $H_0$ compared to the respective non-dynamical case. However, even allowing for a dynamical $w(z)$, when considering the joint analysis of CMB, BAO, and SN data, we are not able to significantly increase $H_0$ to solve the tension, as clear from \autoref{fig:3}.

\item IDE models featuring a dynamical phantom EoS perform worse than the non-dynamical case. We experience the same pattern noticed in the other scenarios: local universe observations rule out the model as a possible solution to the Hubble tension. This becomes pretty much evident when considering CMB, BAO, and SN data altogether, see also \autoref{fig:4}.

\end{itemize}

Overall, our comprehensive reanalysis shows that updated BAO and SN data appear to constrain the possibility that $w_0$IDE or $w(z)$IDE alone can conclusively resolve the Hubble tension for the proposed interaction model. Using the SH0ES calibration for SN (which is a well-motivated choice given the larger values of $H_0$ produced by the CMB-only analysis), we still have room to mitigate, although not fully solve, the Hubble trouble. Additionally, some scatter combinations of data involving BAO also lead to a higher present-day expansion rate for the phantom case. However, considering the joint CMB+SN+BAO measurement, we \textit{always} settle down to values of $H_0$ similar to those inferred within a $\Lambda$CDM-like late-time cosmology. Yet another interesting aspect of our updated analysis is that the state-of-the-art constraints on the DE EoS are relaxed in models that feature a dynamic evolution of $w(z)$ when interactions in the dark sector of the cosmological model are considered. This is due to the increased dimensionality of the parameter space that introduces additional degeneracies among the parameters, leading to larger uncertainties. Therefore, a potential direction for future work could involve extending the analysis to additional observational probes, particularly those related to perturbations and the growth of structures (e.g., weak lensing), which are not accounted for in this work due to the complexity of treating nonlinear scales. Overall, it is worth investigating whether some new ingredients could be added to the IDE cosmology, which can overcome the SN and BAO issues pointed out in this article.

\begin{acknowledgments}
\noindent  We thank the referee for many insightful comments which helped us to improve the quality of the manuscript. SP acknowledges the financial support from the Department of Science and Technology (DST), Govt. of India under the Scheme   ``Fund for Improvement of S\&T Infrastructure (FIST)'' (File No. SR/FST/MS-I/2019/41). EDV is supported by a Royal Society Dorothy Hodgkin Research Fellowship. RCN thanks the financial support from CNPq under the project No. 304306/2022-3, and the FAPERGS for partial financial support under the project No. 23/2551-0000848-3. CvdB is supported (in part) by the Lancaster–Sheffield Consortium for Fundamental Physics under STFC grant: ST/X000621/1. This article is based upon work from COST Action CA21136 Addressing observational tensions in cosmology with systematics and fundamental physics (CosmoVerse) supported by COST (European Cooperation in Science and Technology). We acknowledge IT Services at The University of Sheffield for the provision of services for High Performance Computing.
\end{acknowledgments}
\bibliography{wIDE}
\end{document}